\documentclass[11pt]{article}
\usepackage[T1]{fontenc}
\usepackage{geometry}
\RequirePackage{amsthm,amsmath,amsfonts,amssymb}
\RequirePackage[authoryear]{natbib}
\RequirePackage[colorlinks,citecolor=blue,urlcolor=blue]{hyperref}
\RequirePackage{graphicx}
\RequirePackage{float}
\RequirePackage{bbm}

\usepackage{marvosym}
\usepackage{multirow}

\theoremstyle{remark}

\usepackage[dvipsnames]{xcolor}

\definecolor{lightskyblue}{rgb}{0.53, 0.81, 0.98}
\definecolor{skyblue}{rgb}{0.53, 0.81, 0.92}
\definecolor{deepskyblue}{rgb}{0.0, 0.75, 1.0}
\definecolor{indianred}{rgb}{0.8, 0.36, 0.36}

\usepackage[symbol]{footmisc}

\usepackage{authblk}
\title{Likelihood-based inference for partially observed stochastic epidemics with individual heterogeneity}
\author[1]{Fan Bu\footnote{Now at Department of Human Genetics, University of California, Los Angeles.

Partially supported by DMS-2030355, DMS-2046880, NIH 1R01EB025021.}}
\author[2]{Allison E. Aiello}
\author[1$\dagger\textrm{\Email}$]{Alexander Volfovsky}
\author[1$\dagger$]{Jason Xu}
\affil[1]{Department of Statistical Science, Duke University}
\affil[2]{Gillings School of Global Public Health, University of North Carolina, Chapel Hill}
\affil[$^\dagger$]{Joint last authors; $^\textrm{\Email}$ Corresponding author.}
\date{}

\begin{document}

\maketitle

\begin{abstract}
We develop a stochastic epidemic model progressing over dynamic networks, where infection rates are heterogeneous and may vary with individual-level covariates. 
The joint dynamics are modeled as a continuous-time Markov chain such that disease transmission is constrained by the contact network structure, and network evolution is in turn influenced by individual disease statuses. To accommodate partial epidemic observations commonly seen in real-world data, we propose a likelihood-based inference method based on the stochastic EM algorithm, introducing key innovations that include efficient conditional samplers for imputing missing infection and recovery times which respect the dynamic contact network. Experiments on both synthetic and real datasets demonstrate that our inference method can accurately and efficiently recover model parameters and provide valuable insight at the presence of unobserved disease episodes in epidemic data.  
\end{abstract}

\section{Introduction}
Modern epidemiological studies seek to understand disease dynamics, evaluate intervention strategies and differentiate between population level and individual level effects.
A traditional approach to the modeling of infectious disease relies on 
mechanistic compartmental models, where only the summary of disease statuses of individuals in the population plays a role in understanding the disease dynamics. Examples of such mechanistic compartmental models abound in the epidemiology and mathematical biology literature, e.g. the susceptible-infectious-recovered (SIR) model \citep{kermack1927contribution}. The majority of these simplify disease transmission to a population level event --- as such these models are posed to answer population level questions about disease outbreaks (e.g., ``will the outbreak end?''), but cannot resolve those at the individual level (e.g.,``what is my risk of infection?''). This is exemplified by the ``random mixing'' assumption that underpins many of these models. Under this assumption any infectious individual can transmit the disease to any other susceptible individual with equal chances. However, it is clear that the contact network of individuals plays an integral role in disease transmission and that interventions on individual behavior can change the overall dynamics of an outbreak \citep{eames2003contact,kiss2006infectious,lunz2021quarantine}. 

\begin{figure}[ht]
    \centering
    \includegraphics[trim=0 1in 0 0.7in, clip, width=0.95\textwidth]{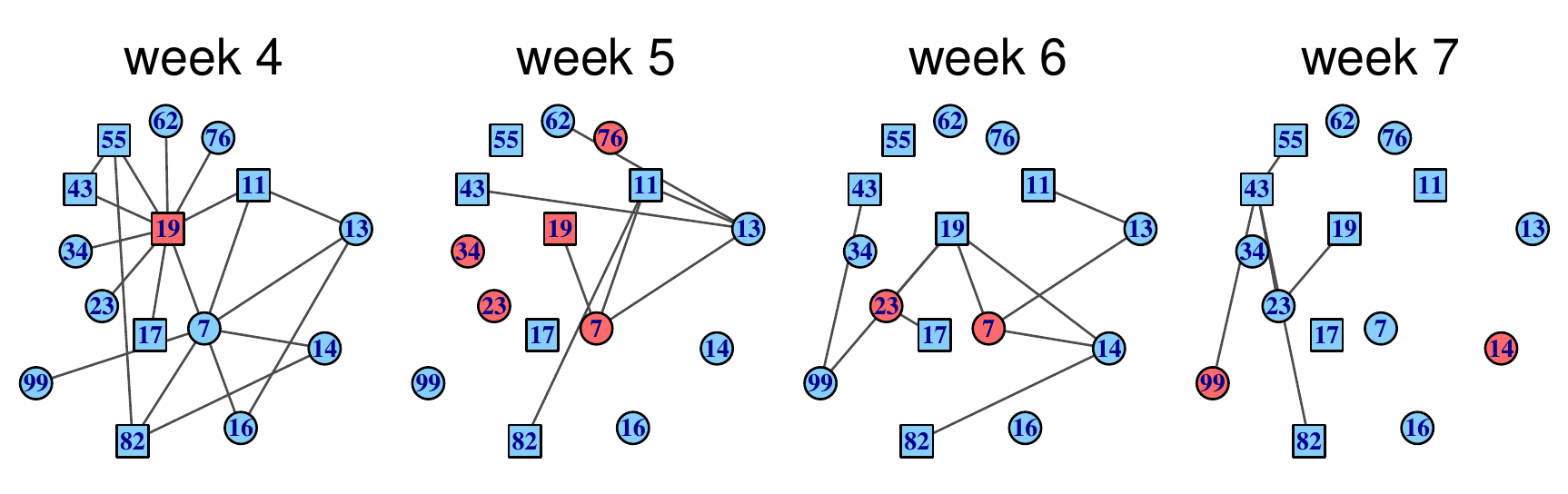}
    \caption[Weekly aggregated contact networks for select participants in the eX-FLU study.]{Weekly aggregated contact networks for select participants in the eX-FLU study. Healthy and sick individuals are marked in {\color{skyblue} \bf blue} and {\color{indianred} \bf red} respectively. Individuals who wash their hands optimally are marked by squares, while those who don't are marked by circles. Here we present the contact networks centered at person 19 from week 4 to week 7 of the study. In week 4, person 19 is infectious, and by week 5 four of his neighbors are infected; later on person 23 and person 7 keep infecting their neighbors and thus spread the disease onto person 99 and 14 by week 7. It is notable that all subsequent infections after 19 are for those who do {not} wash their hands optimally. Moreover, it appears that an individual tends to lose previous contacts after getting sick. }
    \label{fig:subnet-transmission}
\end{figure}

The literature on epidemics diffusing through networks has been greatly bolstered during the SARS-COV-2 pandemic, and a myriad of mechanistic models attempting to capture both the network and disease characteristics have been proposed \citep{ferguson2020report, cencetti2020using,nielsen2021covid,small2020modelling,skums2020global,lee2020engines, soriano2021household}. While a number of these make use of available mobility data collected through powered mobile devices, they still largely operate on the population level with disease dynamics considered at the county or zip-code level. At the same time, high-resolution data collected at the individual level are becoming available, requiring new model development. For example, in Figure~\ref{fig:subnet-transmission} we plot the transmission and interaction dynamics of individuals on a college campus, a snippet from the eX-FLU data we analyze in detail in Section~\ref{sec: data-analysis}.

A number of approaches to individual-level modeling have been introduced in the literature. 
Several papers have introduced a graph-coupled Hidden Markov Model to account for changes in the infectiousness state of individuals \citep{dong2012graph,fan2015hierarchical,fan2016unifying}, and there have been developments in agent-based disease transmission models that consider covariates associated with infections \citep{touloupou2020scalable,ju2021sequential}. Such individual-level inference is very computationally challenging and becomes intractable for even moderately sized datasets with few predictors of interest.  
\cite{bu2020likelihood} introduced an individual level framework for stochastic epidemic processes, where contact information about infections and individual-to-individual interactions are nearly-completely observed. Specifically, the proposed approach introduced an exact sampler when exact recovery times are unobserved that relies on fully observing the infection times, building on
earlier versions of individual-level stochastic models that do not model the network 
\citep{auranen2000transmission,cauchemez2006s,hoti2009outbreaks,britton2010stochastic}. 

We note that 
many of these prior approaches assume that the epidemic process is fully and exactly observed, so inference may proceed via the likelihood of the complete data.
However, available epidemic data typically only provide a partial view on exposure, infection and recovery times --- the exposure time is often ``latent'' because of the incubation or latency period, and infection and recovery times are often not fully kept track of due to limited resolution of data collection or failure of follow-up. 
Even in considerably rich and high-throughput modern datasets such as the eX-FLU study that we analyze in this paper \citep{aiello2016design}, the true exposure times are not observed even though infections can be inferred via daily symptom reporting, and the exact recovery times are not available because only weekly updates were obtained for influenza recoveries. 
Recent work carries out inference from partial epidemic data using the marginal likelihood under simpler SIR or SIS models \citep{ho2018birth,ho2018direct,ju2021sequential}, but the techniques are computationally intensive and difficult to extend to account for the factors (e.g., the contact network structure) we seek to model. Many infectious diseases such as influenza and COVID-19 have a substantial incubation or latency period; our contributions both incorporate a latency period and propose an inference procedure to account for the unknown infection and recovery times through latent variables. 

Lastly, contemporary studies collect information on covariates such as hygiene habits, vaccination statuses, and preventative measures and disease control, which go beyond the simple infection states and contact tracing information that were previously available. The model and procedure of \cite{bu2020likelihood} consider the binary contact status between individuals for modeling infection rates but more flexibility is needed to accommodate heterogeneity in epidemic rate parameters as a function of multiple covariates. 
As a concrete example, consider the role of hygiene habits on disease spread. Previous studies have shown that good hygiene habits such as frequent hand-washing can help reduce the transmission of infectious diseases \citep{aiello2010mask,hubner2013impact,hovi2017intensified,thompson2015healthy}. 
These studies are mostly randomized trials without closed and interacting populations and have thus been analyzed using simple two-sample t-tests or randomization tests \citep{stedman2015outcomes,arbogast2016impact,stedman2015hand,savolainen2012hand}. While this allows us to get a sense of whether hygiene is important overall, it does not quantify the effect of individual hygiene behavior on disease transmission within a joint inference procedure.

In our motivating dataset, the eX-FLU study of influenza-like-illnesses on a college campus \citep{aiello2016design}, raw counts of hand-washers and infection cases suggest that 37\% of those who did not optimally\footnote{In the eX-FLU study ``optimal hand-washers'' were identified through survey questions on the frequency and duration of hand-washing.} wash their hands experienced flu-like-illnesses, while only 24\% of those who did became sick during the study. However, a Fisher's exact test of proportions does not detect this as a significant difference, which is unsurprising as many of the observations are dependent through a network. Inspecting the local network of an infected individual and (manually) tracing the disease transmission over a few weeks (see an illustration in Figure~\ref{fig:subnet-transmission}) suggests that optimal hand-washers are indeed \emph{less likely} to contract the disease, even after contact with infectious individuals. This exploration illustrates how individual-level information affects the complex dependencies between the epidemic process and the contact network in a way that goes undetected in population-level summaries, motivating a framework that directly models individual covariates into the transmission mechanism.


The model proposed in Section~\ref{sec: model-framework}
extends the literature on continuous-time Markov processes to accommodate a contact network, latency period, more missingness in epidemic observations, and individual-level covariates for heterogeneous disease transmission dynamics. We choose to build on the stochastic SEIR model, a variation on the widely used SIR model that explicitly considers the latency period. This flexibility comes at a cost --- exact inference becomes intractable because of the complex dependencies between the covariates, disease process and missingness mechanism. To address this, we derive a stochastic expectation-maximization (stEM) algorithm that can exploit the complete data likelihood while augmenting the data through exact conditional sampling 
\citep{nielsen2000stochastic}. This approach is made computationally tractable due to three key realizations: many of the required computations can be cast as offset Poisson regressions which can be solved efficiently, a rejection sampler for missing exposure times can be deployed naively in parallel, and an exact sampler for missing recovery times can be adapted from \cite{bu2020likelihood}. We further leverage the theoretical guarantees for stEM to compute conservative asymptotic variance estimates.

The remainder of this paper is organized as follows: In the next section, we introduce our model framework. Sections~\ref{sec: inference} and \ref{sec: sto-EM} discuss our proposed inference procedures for complete data and partial data. We evaluate the performance of the proposed inference methods through simulation experiments in Section~\ref{sec: simulation-experiments}, and finally apply our model to analyzing the eX-FLU dataset in Section~\ref{sec: case-study}. 

\section{Model framework}
\label{sec: model-framework}
We adopt a stochastic compartmental model for epidemics, where all members of the target population are divided into non-overlapping subsets related to their disease statuses, and the mechanism of disease spread is described by the transition between disease statuses for each individual. We base our epidemic model on the SEIR model with four disease statuses: $S$ (susceptible), $E$ (exposed), $I$ (infectious), and $R$ (recovered or removed). An $S$ individual may get exposed (and thus become an $E$ person) upon contact with an $I$ individual, and an infectious ($I$) person will eventually recover and transition to the $R$ status. In this model, the $E$ status resembles a latency period and does not entail any transmissibility, and a recovered person enjoys immunity to the disease and therefore no longer contributes to the contagion process. 

These disease spread dynamics evolve as a continuous-time Markov chain (CTMC) defined through exponentially distributed waiting times between consecutive events \citep{guttorp2018stochastic}. This implies that the disease process progresses as a series of competing Poisson processes at the individual level. For example, suppose $\beta_{ij}$ is the rate of exposure between an infectious person $i$ and an susceptible person $j$ who are in contact at time $t$. Then the probability of $j$ getting exposed (thus becoming an $E$ person) at time $t+h$ for small $h>0$  is
\begin{equation}
    Pr(j \text{ gets exposed by } i \text{ by } t+h \mid i, j \text{ in contact at }t) = \beta_{ij} h + o(h).
\end{equation}

Given that the contact structure of the population is subject to change in time as well, we extend the CTMC model to the dynamics of the contact network.  For any pair of individuals $i$ and $j$, they either share an undirected contact link (``connected'') or they do not (``disconnected'');  the contact network can thus be represented by a binary symmetric matrix $W$ called the adjacency matrix. 
Its dynamics are described at the pairwise level, where each entry $W_{ij}$ evolves as a CTMC that takes values in $\{0,1\}$. For example, if individuals $i$ and $j$ are disconnected at time $t$, with link creation rate $\alpha_{ij}$, the probability of them engaging in contact by time $t+h$ for small $h>0$ is
\begin{equation}
    Pr(W_{ij} = 1 \text{ at time } t+h \mid W_{ij} = 0 \text{ at time } t) = \alpha_{ij} h + o(h).
\end{equation}

\begin{figure}[ht]
    \centering
    \includegraphics[width=0.9\textwidth,page=4,trim={0 1.8in 0 1.8in},clip]{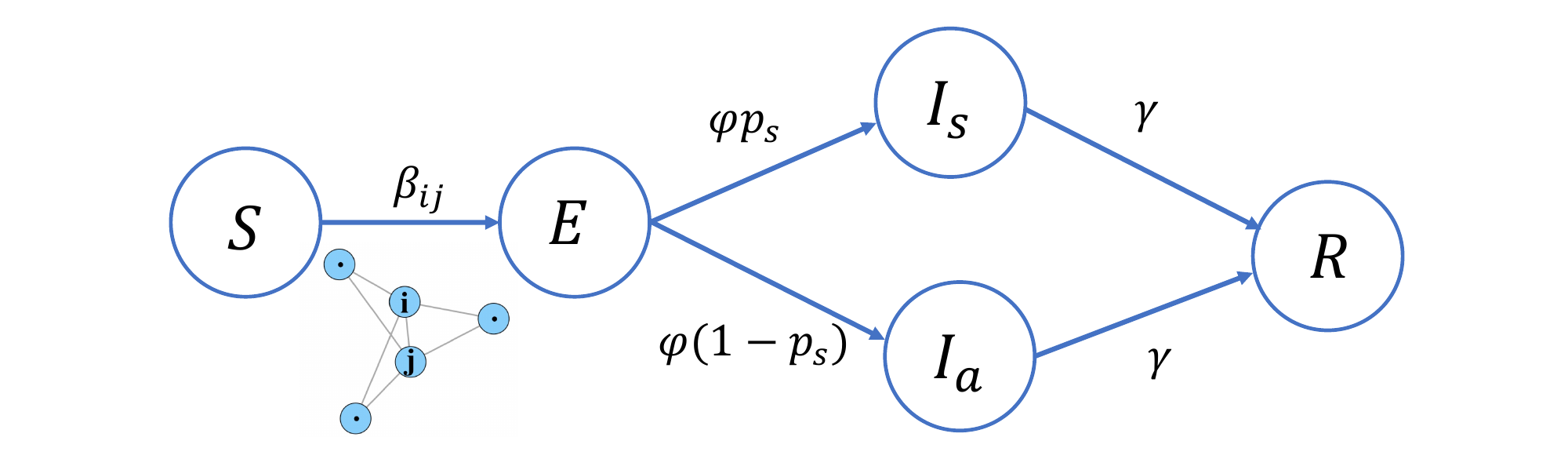}
    \caption{Diagram of the epidemic process: an extension of the stochastic SEIR model, with heterogeneous exposure rates and two sub-types of infectives. Disease transmission (exposure) is conditioned on pairwise contact status in the dynamic contact network.}
    \label{fig:SEIIR-diagram}
\end{figure}

We will consider heterogeneous exposure rates, allowing individual characteristics and network information to play a substantive role in transmission probabilities. Moreover, we consider two types of infectious individuals --- these can be thought of as symptomatic and asymptomatic cases
who exert different transmission forces on the population. This is summarized in Figure~\ref{fig:SEIIR-diagram}.


\noindent\textbf{Model Specification. } 
Our goal is to construct an individualized framework that can capture the \emph{interplay} between the epidemic process and the evolution of the contact network. Disease transmission relies on contact between individuals, while the change in contacts, in turn, depends on individual disease statuses. To do so, we model the joint evolution of the epidemic and network processes as the combination of continuous-time Markov chains for individuals (or pairs of individuals) in the population. 
At any time point $t$ in the process, conditioned on the current status of the process $\mathcal{Z}_t$, five types of events may occur for an individual or a pair of individuals: \emph{exposure} (an $S$ person becomes exposed by an $I$ person), \emph{manifestation} (an $E$ person becomes infectious after a latency period), \emph{recovery} (an $I$ person recovers and becomes a $R$ person), \emph{link activation} (a previously disconnected pair get connected in the network), and \emph{link termination} (a previously connected pair break off their contact). 

We accommodate different types of individual heterogeneity in the disease transmission dynamics: (1) people may exhibit different levels of susceptibility that can be explained by individual characteristics such as health conditions, hygiene habits, and behavioral choices; (2) those who are infectious might not be equally contagious to the susceptible population (e.g., symptomatic and asymptomatic cases for COVID-19); (3) contact rates in the network may vary in time to reflect phases of social intervention and/or behavioral changes as response to an epidemic (e.g., a pre- and post-lockdown period denoted by $\mathcal{T}_0$ and $\mathcal{T}_1$, respectively);
(4) contact rates in the network may vary between pairs of individuals based on their healthy (denoted by $H$, the collection of $S$, $E$, and $R$ statuses) versus infectious (denoted by $I$) status. 


Combining the above descriptions, we design the model framework as follows. For any members $i$ and $j$ in the target population, given the current system state $\mathcal{Z}_t$ at time $t > 0$, one of the following five events may occur after an exponentially distributed waiting time with the associated rate (see Figure~\ref{fig:SEIIR-diagram} for a summary of the epidemic process):
\begin{itemize}
    \item \textbf{Exposure}. If $i$ is infectious, $j$ is susceptible, and they are in contact at time $t$, then $i$ exposes $j$ with instantaneous rate
    $\beta_{ijt}$ that can be decomposed as
    \begin{equation}
        \label{eq:beta-ij}
    \log\beta_{ijt} = \log \beta + \eta_i(t) + b_S^T x_j,
    \end{equation}
    where $\beta$ is the baseline exposure rate, $\eta_i(t)$ represents $i$'s contagiousness level at time $t$ (details in the next item), and $b_S$ are the regression coefficients on $j$'s individual characteristics $x_j$ that account for additional heterogeneous effects of susceptibility. (For example, $x_j$ can characterize if person $j$ has been vaccinated, washes their hands frequently and/or wears a mask during the flu season, which may reduce $j$'s exposure risk.)
    \item \textbf{Manifestation}. If $i$ (or $j$) is exposed, then he or she becomes infectious with rate $\varphi$. For additional generality we assume that any infective (status $I$ person) gets assigned to one of two categories with different levels of contagiousness: $I_s$ (``symptomatic'') or $I_a$ (``asymptomatic'' or ``less symptomatic''). Individuals are assigned to the first category with probability $p_s$ and to the second with probability $(1-p_s)$.\footnote{Note that it is straightfoward to introduce more sub-types of infectives or include continuous and even time-varying explanatory variables for the function $\eta_i(t)$, if more intricate modeling of heterogeneous transmissibility is necessary.} With the two-type infective setup, the $\eta_i(t)$ term in Eq.~\eqref{eq:beta-ij} can be written as
    \begin{equation}
        \label{eq: eta-function}
        \eta_i(t) = \eta \mathbbm{1}(i \text{ is } I_s \text{ at } t),
    \end{equation}
    which means that an $I_s$ person is on average $e^{\eta}$ times more infectious than an $I_a$ individual.
    \item \textbf{Recovery}. If $i$ (or $j$) is infectious, then he or she recovers with rate $\gamma$.
     \item \textbf{Link activation}. If $i$ and $j$ are not in contact, then they get into contact with rate $\alpha_{ijt}$, where  
    \begin{equation}
    \label{eq:alpha-ij}
    \alpha_{ijt} = \alpha_{A_{it}A_{jt}0}\mathbbm{1}(t \in \mathcal{T}_0) +\alpha_{A_{it}A_{jt}1}\mathbbm{1}(t \in \mathcal{T}_1),
    \end{equation}
    where $A_{it}$ is the healthy or infectious status of person $i$ at time $t$ and $\alpha_{ABk}$ stands for the activation rate of link type $A-B$ in phase $\mathcal{T}_k$ ($A,B\in \{H,I\}$ and $k\in \{0,1\}$).
    \item \textbf{Link termination}. If $i$ and $j$ are in contact, then they break off the contact with rate $\omega_{ijt}$, where 
    \begin{equation}
    \label{eq:omega-ij}
    \omega_{ijt} = \omega_{A_{it}A_{jt}0}\mathbbm{1}(t \in \mathcal{T}_0) +\omega_{A_{it}A_{jt}1}\mathbbm{1}(t \in \mathcal{T}_1),
    \end{equation}
    and $\omega_{ABk}$ stands for the termination rate of link type $A-B$ in phase $\mathcal{T}_k$.
\end{itemize}

Here $\mathcal{T}_0$ and $\mathcal{T}_1$ represent two time phases of different social behaviors that partition the entire observation time window $(0,T)$ (e.g., $\mathcal{T}_0$ and $\mathcal{T}_1$ can be intermittent lockdown and no-lockdown phases).

The link rates are dependent on the individual disease statuses $H$ or $I$ since we wish to characterize the social adaptation behavior in response of an epidemic -- for instance, a healthy-infectious ($H-I$) pair that are in contact might be more likely to disconnect from each other than a pair of healthy individuals to avoid disease transmission.
Further, since we assume the contact network is symmetric and only dependent on healthy or infectious statuses, the link rates satisfy $\alpha_{HI_k} = \alpha_{IH_k}$ and $\omega_{HI_k} = \omega_{IH_k}$ for $k=0,1$.

Note that we can also include individual-level covariates in the link activation and termination rates ($\alpha_{ijt}$ and $\omega_{ijt}$) to allow for more heterogeneity in the network dynamics, but choose to focus on individual variability in the epidemic process in the main text, and relegate details of heterogeneous network link rates to Section S1 of the Supplementary Material \citep{bu2021supplement}.

\section{Inference}
\label{sec: inference}
In this section, we demonstrate how to learn model parameters in the missing data setting. Though our focus will be on inference in the partially observed regime, we begin by describing key inferential terms related to the complete data setting as these quantities will play a role in our stochastic EM algorithm. 

\subsection{Inference with complete data}
\label{sec: inference-complete}
A complete dataset refers to the fullly observed event sequence between time $0$ and maximum time $T$ ($>0$) of one realization from the generative model. In particular, if we were to continuously observe an epidemic as it progresses under this model, we would have access to (1) exact times of exposure ($t_i^{(E)}$), manifestation ($t_i^{(I)}$), recovery, and link activation and termination; (2) the identities of the individuals involved in each event; (3) the $I_s$ or $I_a$ subtype allocation for each infectious individual at the time of their manifestation event; and (4) the contact network structure as well as initial disease statuses of all individuals at time $0$. 


Given the complete data (or equivalently, sufficient statistics summarizing the data) and all individual characteristics $\{x_i\}$, we can write down the complete data likelihood with respect to the parameters of the model $\Theta = \{\beta, \varphi, \gamma, \eta, b_S, \boldsymbol\alpha, \boldsymbol\omega\}$. Here the vectors denote $\boldsymbol\alpha = \{\alpha_{ABk}\}_{k\in\{0,1\}, (A,B)\in \mathcal{S}}$, $ \boldsymbol\omega = \{\omega_{ABk}\}_{k\in\{0,1\}, (A,B)\in \mathcal{S}}$, and we index by $\mathcal{S}=\{(H,H), (H,I), (I,I)\}$ the set of all pair types. The likelihood takes the form
\begin{align}
    & L(\Theta; \text{complete data})  \label{eq: complete-data-likelihood}\\
   &= \beta^{n_E} \gamma^{n_R} \varphi^{n_I} p_s^{n_{I_s}} (1-p_s)^{n_{I_a}} \prod_{i: i \text{ got exposed}} e^{b_S^T x_i} \left[I^{a}_i(t_i^{(E)})+I^{s}_i(t_i^{(E)}) e^{\eta}\right] \nonumber \\
    \times & \prod_{k=0,1}\prod_{(A,B)\in \mathcal{S}} \left[\left(\alpha_{ABk}\right)^{C_{ABk}} \left(\omega_{ABk}\right)^{D_{ABk}}\right] 
    \nonumber \\
    \times & \exp\left(-\int_0^T \left[\beta \sum_{i=1}^N e^{b_S^T x_i} \left[I^{a}_i(t)+I^{s}_i(t) e^{\eta}\right]\mathbbm{1}(i\text{ is susceptible at } t) + \gamma (I^a(t)+I^s(t)) + \varphi E(t)\right] dt \right)  \nonumber \\
    \times & 
    \exp\left( -\int_0^T  \sum_{k=0,1}\sum_{(A,B)\in \mathcal{S}} \left[ \alpha_{ABk} M^d_{AB}(t) + \omega_{ABk} M^c_{AB}(t) \right] \mathbbm{1}(t \in \mathcal{T}_k)dt \right). \nonumber
\end{align}
Table~\ref{tab:notation} summarizes the notation used in the expressions above. 

Since the generative model is a CTMC comprised of individual-level Poisson processes, the above likelihood can be decomposed into epidemic-related components (1st and 3rd lines above) and network-related components (2nd and 4th lines), where each component is simply the product of all the exponential rates for 
inter-event times. 
Some of those inter-event rates can be considered at the population level, and so only require bookkeeping of aggregated counts (e.g., $n_E =$ total number of exposed cases , $E(t) =$ number of $E$ people at time $t$, and $C_{HI0} =$ number of link activation events for $H$-$I$ pairs in phase $\mathcal{T}_0$). 
Other inter-event rates, however, depend on individual-level information, and so require tracking terms such as the time-varying neighborhood structure and the exposure and manifestation times for each individual. Note that, for consistency, we set individual $i$'s exposure time $t_i^{(E)}$ and manifestation time $t_i^{(I)}$ to $T$ if $i$ has never been exposed or manifested.

\begin{table}[H]
    \caption{Explanation of notation}
    \label{tab:notation}
    \small
    \begin{tabular}{@{}ll@{}}
    \hline
    Notation &  Explanation\\
    \hline
    $N$ & total population size (assumed fixed)\\ [4pt]
    $n_{I_s}, n_{I_a}, n_E, n_I, n_R$ & total number of $I_s$, $I_a$, exposed ($E$), infectious ($I$) and recovered ($R$) cases\\[4pt]
    $I^a_i(t), I^s_i(t)$ & total number of $I_a$ and $I_s$ neighbors of $i$ at time $t$\\ [4pt]
    $I^a(t), I^s(t), E(t)$ & total number of status $I_a$, $I_s$, and $E$ individuals in the population at time $t$\\ [4pt]
    $t_i^{(E)}, t_i^{(I)}$ & exposure time and manifestation time for individual $i$ (set to $T$ if never exposed/manifested) \\ [4pt]
    $C_{ABk}, D_{ABk}$ & total number of link activation \& termination events among type $A-B$ pairs in phase $\mathcal{T}_k$\\ [4pt]
    $M^c_{AB}(t), M^d_{AB}(t)$ & number of connected \& disconnected type $A-B$ pairs at time $t$\\ [2pt]
    \hline
    \end{tabular}
\end{table}


Despite a lengthy likelihood expression, parameter estimation is straightforward when complete data are available. We can obtain closed-form maximum likelihood estimates (MLEs) for most of the parameters, and find the remaining MLEs for parameters $\beta, \eta$ and $b_S$ through a simple numerical procedure. This suggests that likelihood-based inference given completely observed data is easily implementable through a few lines of code, and can be modularized toward inference when some data are missing (discussed in the next section). 

The MLEs of all parameters can be obtained by taking partial derivatives of the log-likelihood (denoted by $\ell$) as follows and setting them all to zero: 
\begin{eqnarray}
 \label{eq:beta-derivative}
    \frac{\partial \ell}{\partial \beta} &=& \frac{n_E}{\beta} - \sum_{i=1}^N e^{x_i^T b_S} \left[\int_0^{t_i^{(E)}} (I^a_i(t) +I^s_i(t) e^{\eta}) dt\right],\\
    \label{eq:b-S-derivative}
    \frac{\partial \ell}{\partial b_S} &=& \sum_{i: i\text{ got exposed}} x_i - \beta\sum_{i=1}^N e^{x_i^T b_S} \left[\int_0^{t_i^{(E)}} (I^a_i(t) +I^s_i(t) e^{\eta}) dt\right] x_i,\\
    \label{eq:Eeta-derivative}
    \frac{\partial \ell}{\partial e^{\eta}} &=& \sum_{i: i\text{ got  exposed}}\frac{I^s_i(t_i^{(E)})}{I^a_i(t_i^{(E)}) +I^s_i(t_i^{(E)}) e^{\eta}} - \beta \sum_{i=1}^N e^{x_i^T b_S} \int_0^{t_i^{(E)}} I^s_i(t) dt,\\
    \label{eq:varphi-derivative}
    \frac{\partial \ell}{\partial \varphi} &=& \frac{n_I}{\varphi} - \int_0^T E(t) dt,\\
    \label{eq:ps-derivative}
    \frac{\partial \ell}{\partial p_s} &=& \frac{n_{I_s}}{p_s} - \frac{n_{I_a}}{1-p_s},\\
    \label{eq:gamma-derivative}
    \frac{\partial \ell}{\partial \gamma} &=& \frac{n_R}{\gamma} - \int_0^T I(t) dt,\\
    \label{eq:alpha-derivative}
    \frac{\partial \ell}{\partial \alpha_{ABk}} &=& \frac{C_{ABk}}{\alpha_{ABk}} - \int_{\mathcal{T}_k} M^d_{AB}(t) dt \quad (\text{for }k=0,1, (A,B)\in\mathcal{S}),\\
    \label{eq:omega-derivative}
    \frac{\partial \ell}{\partial \omega_{ABk}} &=& \frac{D_{ABk}}{\omega_{ABk}} - \int_{\mathcal{T}_k} M^c_{AB}(t) dt \quad (\text{for }k=0,1, (A,B)\in\mathcal{S})
    .
\end{eqnarray}

The MLEs for parameters $\varphi,\ p_s,\ \gamma$ and $\alpha_{ABk},\ \omega_{ABk}$ ($k=0,1, (A,B)\in\mathcal{S}$) have closed-form expressions
\begin{eqnarray}
\hat\varphi &=& \frac{n_I}{\int_0^T E(t) dt},  \quad
\hat{p}_s = \frac{n_{I_s}}{n_I}, \quad
\hat\gamma = \frac{n_R}{\int_0^T I(t) dt}, \nonumber\\
\hat\alpha_{ABk} &=& \frac{C_{ABk}}{\int_{\mathcal{T}_k} M^d_{AB}(t) dt}, \quad
\hat\omega_{ABk} = \frac{D_{ABk}}{\int_{\mathcal{T}_k} M^c_{AB}(t) dt}, \nonumber
\end{eqnarray}
while the MLEs for $\beta$, $\eta$ and $b_S$ can be solved from equations~\eqref{eq:beta-derivative}-\eqref{eq:Eeta-derivative} numerically. An efficient iterative procedure is detailed in Section S2 of the Supplementary Material \citep{bu2021supplement}; in particular, steps for solving the MLE of $b_S$ can be largely simplified based on the observation that it is equivalent to solving for the linear coefficient of a Poisson regression model with individual offset. 


\subsection{Inference with partial epidemic observations}
\label{sec: inference-partial} 

As demonstrated in the previous subsection, parameter estimation is relatively straightforward when the full event sequence is observed. However, as discussed in the introduction, real-world epidemic data rarely rarely include measurements of each epidemic event. 
In the case of the eX-FLU study, true exposure times are not available even though the data contain daily symptom reports. This is because there is typically an incubation period for people who contract the flu. Similarly, exact recovery times are not available in the eX-FLU study, with recoveries discernible only at a weekly resolution from epidemic surveys. Therefore, we need to consider inference with partially observed epidemic data, in particular with exposure times and recovery times unknown. 

%
To this end, we derive a method based on the stochastic expectation-maximization (stEM) algorithm \citep{celeux1985sem}. Expectation maximization (EM) offers an approach to efficiently carry out maximum likelihood estimation for continuous-time Markov chain models in missing data settings \citep{doss2013fitting,xu2015likelihood,guttorp2018stochastic}. Imputing the missing data in the E-step requires access to the conditional expectation, and stEM is a variant that builds an approximation to the conditional expectation using augmented data obtained via conditional simulation. To be more precise, let $X$ denote the observed data and $Z$ be the missing data; a general outline of the stochastic EM algorithm for estimating $\theta$ is as follows:

For $s=1:\text{maxIter}$, do
\begin{itemize}
    \item (E-step) draw one sample of missing data, $Z^{(s)}$ from its conditional distribution $p(Z \mid X, \theta^{(s-1)})$, and then let 
    \begin{equation*}
        Q(\theta \mid \theta^{(s-1)}) = \log L(\theta ; X, Z^{(s)});
    \end{equation*}
    \item (M-step) maximize with respect to target function $Q(\theta \mid \theta^{(s)})$ to update $\theta$:
    \begin{equation*}
        \theta^{(s)} = \arg\max_{\theta}Q(\theta \mid 
        \theta^{(s-1)}).
    \end{equation*}
\end{itemize}

There are two advantages of this approach. 
First, in the E-step, integrating to obtain an expected log-likelihood (as in the traditional EM algorithm) is replaced by sampling, which avoids the often intractable marginalization step in the case of complex models \citep{renshaw2015stochastic,xu2015efficient,stutz2021}.
Second, the M-step simply requires solving for the MLEs given a version of the complete data, which is often straightforward, and has been discussed in the previous subsection for the present setting. 

These advantages come at the cost of a potential challenge: we have to conditionally sample the missing data given our observed data and current parameter estimates. In our framework, this is equivalent to sampling event times of a continuous-time Markov chain conditioned on end-points, a notably difficult problem \citep{hobolth2009simulation,rao2013fast}. The following section will focus on the conditional sampling (or ``imputing'') of missing data, i.e., missing exposure and recovery times. 

\section{Stochastic EM approach}
\label{sec: sto-EM}
Let $\mathbf{t^{(E)}}$ and $\mathbf{t^{(R)}}$ denote all  missing exposure times and recovery times, respectively. We assume that all  manifestation times $\{t_i^{(I)}\}$ are observed,\kern-.1em\footnote{This assumption is reasonable as manifestation times can be collected via daily symptom monitoring or frequent screening or testing.} and there is no missingness in the contact network events given high-resolution contact-tracing. Thus, our inference procedure with partial epidemic observations can be outlined as follows.

For $s=1:\text{maxIter}$, do
\begin{enumerate}
    \item sample missing exposure times $\mathbf{t^{(E)}}^{(s)}$ from their joint conditional distribution $p(\mathbf{t^{(E)}} \mid \text{observed events}, \mathbf{t^{(R)}}^{(s-1)}, \Theta^{(s-1)})$;
    \item sample missing recovery times $\mathbf{t^{(R)}}^{(s)}$ from their joint conditional distribution $p(\mathbf{t^{(R)}} \mid \text{observed events}, \mathbf{t^{(E)}}^{(s)}, \Theta^{(s-1)})$;
    \item combine the sampled event times in Steps 1 and 2 with observed data to form an augmented dataset, then solve for the MLEs with the augmented dataset to get updated parameter estimates $\Theta^{(s)}$.
\end{enumerate}

Since Step 3 is already addressed in Section 3.1, we now address Steps 1 and 2 separately. 

\subsection{Step 1: conditional sampling of missing exposure times}
\label{sec: missing-expo-time}


Inspecting the complete data likelihood in Eq.~\eqref{eq: complete-data-likelihood} reveals that given all the other event times, person $i$'s exposure time $t$ is independent from other individuals' exposure times, and thus the joint conditional density for $\mathbf{t^{(E)}}$ can be factorized into individual density components of exposure. Thus, it suffices to derive the conditional density for $i$'s exposure time $t_i^{(E)}$, which is assumed to lie within a ``plausible'' latency interval $L_i = (t_{\min}^{i}, t_{\max}^{i})$. Note it is always valid to choose this interval as $L_i = (0, t_i^{(I)})$, meaning that  $i$'s exposure time may occur any time after time $0$ and before $i$'s manifestation time. In practice, one may alternatively specify a shorter plausible interval based on prior knowledge about latency duration, which can improve computational efficiency.\footnote{For example, if we believe that latency should be longer than 2 days but shorter than 2 weeks, then we may set $t_{\min}^{i} = \max(0, t_{i}^{(I)} - 14)$ and $t_{\max}^{i} = \max(0, t_{i}^{(I)} - 2)$.}
Then, $i$'s instantaneous risk function of exposure and contracting the disease during $L_i$ can be written as (here let $\delta_i = b_S^T x_i$)
\begin{equation}
    \label{eq: infection-risk}
    \lambda_{i}(t) = \beta e^{
    \delta_i}(I^a_i(t) + e^{\eta} I^s_i(t)), \quad t\in L_i.
\end{equation}
This exposure hazard is a \emph{step function} with change points occurring when either (1) $i$ activates a link with an $I_s$ or $I_a$ person, (2) $i$ deactivates a link with an $I_s$ or $I_a$ person, (3) one of $i$'s contacts enters status $I_s$ or $I_a$, or (4) one of $i$'s contacts exits status $I_s$ or $I_a$. For person $i$, denote this set of change points $\{t_{j}\}_{j=0}^{n_i}$, with $t_0 = t_{\min}^i, t_{n_i} = t_{\max}^i$\footnote{Here we suppress the notation that $t_j$ is associated with individual $i$ to avoid double subscripts, focusing on one individual during exposition.}. This set defines a partition 
$\mathcal{A}_i$ of the latency interval $L_i$ such that on each sub-interval $A_j = (t_{j-1}, t_{j})\in \mathcal{A}_i$, $\lambda_{i}(t) \equiv \lambda_{j}$ is constant.

Thus, the conditional density for $i$'s exposure time $t$ 
can be expressed as
\begin{align}
    &p_i(t \mid t_i^{(I)}, \beta, \delta_i, \eta, \varphi, \text{network events}) \nonumber\\
    \label{eq: exposure-density-original}
     =& \frac{\lambda_{i}(t) \exp\left(-\int_{t_{\min}^{i}}^t \lambda_{i}(u)du\right) \times \varphi \exp(- \varphi (t_i^{(I)} - t)) \mathbbm{1}(t_{\min}^{i} < t < t_{\max}^{i})}{C_i(t_{\min}^{i},t_{\max}^{i})},
\end{align}
where the normalizing constant $C_i(t_{\min}^{i},t_{\max}^{i})$ can be explicitly evaluated since $\lambda_i(t)$ is a step function. 


We derive a rejection sampler for the missing exposure time $t_i^{(E)}$ from $p_i(t)$, with the ``plausible interval'' $L_i$ set as $(0, t_i^{(I)})$. 
 Consider the following proposal density 
\begin{equation}
\label{eq:proposal-density}
    q_i(t) = \frac{\lambda_{i}(t) \exp\left(-\int_{0}^t \lambda_{i}(u)du\right)\mathbbm{1}(0 < t < t_i^{(I)})}{1-\exp\left(-\int_{0}^{t_i^{(I)}} \lambda_{i}(u)du\right)},
\end{equation}
which is the density function of a truncated inhomogeneous Exponential distribution with rate $\lambda_{i}(t)$. A rejection sampler for $t_i^{(E)}$ runs in two steps:
\begin{enumerate}
    \item Sample $t$ from $q_i(t)$, an inhomogeneous Exponential with rate $\lambda_{i}(t)$ truncated on $L_i$:
    \begin{enumerate}
        \item sample an interval $A_j$ (recall that the risk function is constant on each interval) via
        \begin{align}
            \text{Pr}(t \in A_j \mid \lambda_{i}(t)) 
            =& \frac{\exp\left(-\sum_{k<j} \lambda_k \text{len}(A_k) \right)-\exp\left(-\sum_{k\leq j} \lambda_k \text{len}(A_k)\right)}{1-\exp\left(\sum_{k=1}^{n_i} \lambda_k \text{len}(A_k)\right)}, \nonumber
        \end{align}
        where we denote the length of $A_j$ by len($A_j$).
        \item within interval $A_j$, sample $t \sim Exp(\lambda_j)$ truncated on interval $(t_{j-1}, t_{j})$.
    \end{enumerate}
    \item Compute the acceptance probability for $t$ by (here $M>1$ is a constant)
    \begin{equation}
        \frac{p_i(t)}{M q_i(t)} = \exp(- \varphi (t_i^{(I)} - t)),
    \end{equation}
    and draw $U \sim Unif(0,1)$; accept $t$ as a sample of $t_i^{(E)}$ if $U < \exp(- \varphi (t_i^{(I)} - t))$, and otherwise go back to Step 1 and repeat. 
\end{enumerate}

A full derivation of the above (importantly showing that $M > 1$) is provided in Section S3 of the Supplementary Material \citep{bu2021supplement}. 
Note that it is easy to generalize to a setting with other choices of the plausible interval bounds $t_{\min}^i$ and $t_{\max}^i$, which simply entails changing the limits of integration in the above derivations. 

Step 1 of the stEM procedure can be carried out by running the rejection sampler above, and we may further speed up computation by running the sampler for each person $i$ \emph{in parallel}. In the simulations in Section~\ref{sec:sim-partial}, the rejection sampler accepts approximately 45\% of proposals. 


\subsection{Step 2: conditional sampling of missing recovery times}
Every infectious individual recovers with rate $\gamma$ independently of other members in the population, but when conditionally sampling missing recovery times, we have to make sure that the imputed timepoints are compatible with observed data and the sampled exposure times \citep{cauchemez2008likelihood,fintzi2017efficient}. The conditional samples of missing recovery times should satisfy two conditions: first of all, an individual $q$ cannot recover before a time $t$ if $q$ is known to be still infectious by $t$; and more importantly, if another individual $p$ gets exposed during his contact with $q$, then the recovery time for $q$ cannot leave $p$ with no possible infection source. 

Sampling missing recovery times, therefore, amounts to conditionally sampling event times with endpoints restricted by count and contact data. 
This challenging task was addressed by the DARCI algorithm developed in \cite{bu2020likelihood} (Proposition 4.2) for a simpler epidemic model with only one type of infectives. Here, we cannot directly use DARCI because the different transmissibility levels of $I_a$ and $I_s$ individuals must be taken into account when we consider the possible ranges of recovery times to ensure the existence of viable infection sources for those exposed. Instead, we adapt the DARCI algorithm to accommodate our two types of infectives. Below we discuss the details of the modified DARCI procedure. 

Utilizing the Markov property, this algorithm first segments the observation window $(0, T]$ into contiguous, non-overlapping time intervals. On each time interval $(u,v]$, the disease statuses of all people are known at endpoints $u$ and $v$, and thus we know the set of people $\mathcal{Q}$ who should recover during $(u,v]$.\footnote{For example, in the eX-FLU study, weekly epidemic surveys would provide information on if someone felt sick during each week.} Further, conditioned on the sampled exposure times, we also know the infection/exposure cases and their exposure times during $(u,v]$ and let these individuals be $\mathcal{P}$. We sample the missing recovery times $\{t_q^{(R)}\}_{q\in \mathcal{Q}}$ in the following steps:
\begin{enumerate}
    \item Initialize a vector of ``feasible lower-bounds'' $\text{LB}$ of length $|\mathcal{Q}|$ with $\text{LB}_q = u$ for every $q \in \mathcal{Q}$; for any $p \in \mathcal{P} \cap \mathcal{Q}$, further set $\text{LB}_p = i_p$, where $i_p$ is $p$'s exposure time;
    \item Arrange the set of exposed individuals $\mathcal{P}$ in the order of $\{p_1, p_2, \ldots, p_{|\mathcal{P}|}\}$ such that their exposure times $i_{p_1} < i_{p_2} < \ldots < i_{p_{|\mathcal{P}|}}$, and for each $p \in \mathcal{P}$ (chosen in the arranged order), examine $p$'s ``potential infectious neighborhood''
    \begin{equation*}
            \mathcal{I}_p = \mathcal{N}_p(i_p) \cap \left(\mathcal{I}(i_p) \cup \mathcal{Q}\right),
    \end{equation*}
    where $\mathcal{N}_p(t)$ is the set of $p$'s neighbors at time $t$, and $\mathcal{I}(t)$ is the set of \emph{known} infectious individuals at time $t$ 
    .\\ 
    If $\mathcal{I}_p \subset \mathcal{Q}$ (i.e., potential infection sources are all members of $\mathcal{Q}$), then select one $q \in \mathcal{I}_p$, with probability
    \begin{equation}
    \frac{\mathbbm{1}(q \text{ is } I_a \text{ at } i_p) + e^{\eta} \mathbbm{1}(q \text{ is } I_s \text{ at } i_p)}{ \sum_{q' \in \mathcal{I}_p} \mathbbm{1}(q' \text{ is } I_a \text{ at } i_p) + e^{\eta} \mathbbm{1}(q' \text{ is } I_s \text{ at } i_p)},
    \end{equation}
    and set $\text{LB}_q = i_p$.
    \item Draw recovery times $t_q^{(R)}  \stackrel{ind}{\sim} \text{TEXP}(\gamma, \text{LB}_q , v)$, where $\text{TEXP}(\gamma, s, t)$ is a truncated Exponential distribution with rate $\gamma$ and truncated on the interval $(s,t)$.
\end{enumerate}

Despite the dense notation, the intuition of this sampling algorithm is straightforward: if person $q$ is the \emph{only} possible infection source of person $p$, then $q$ should wait until $p$ gets exposed before he or she recovers, so that the resulting augmented data are consistent with the observed data. 

We note that the conditional sampling of recovery times is parallelizable as well, since operations on each interval $(u,v]$ are independent and thus can be run in parallel. 

\subsection{Uncertainty quantification and improving efficiency via averaging}
\label{sec: variance-estimation}
While the algorithm proposed above provides a way to estimate $\Theta$, we may
further quantify uncertainty in our estimates by leveraging expressions for their asymptotic variances 
by appealing to results established in \cite{nielsen2000stochastic}.

Let $\hat{\Theta}$ denote the parameter estimates from the stochastic EM algorithm, 
and $\Theta_0$ be the true parameter values. Then the asymptotic variance matrix of $\hat{\Theta}$ is
\begin{equation}
    \label{eq: vanilla-variance}
    I(\Theta_0)^{-1} + I(\Theta_0)^{-1}[I_p - (I_p + F(\Theta_0))^{-1}],
\end{equation}
where $I(\Theta_0)$ is the information matrix of the observed data 
evaluated at the true parameter values, $F(\Theta_0)$ is a matrix representing the fraction of missing information due to partial observations, and $I_p$ is the $p$-dimensional identity matrix. 
We may decompose Eq.~\eqref{eq: vanilla-variance} into two parts: the first term is the asymptotic variance of the observed data MLE, while the second term accounts for the additional variance arising from stochastic simulations (i.e., the conditional sampling). It can be shown that the latter contribution increases the variance by no more than 50\% over the observed data MLE (Proposition 4 in \cite{nielsen2000stochastic}). 

Moreover, the first term in the asymptotic variance formula can be regarded as a constant given a model and a fixed sample size, but the second term can be reduced by averaging over either across iterations in one run, or over multiple runs of the algorithm. That is, to improve efficiency, we may adopt two strategies:
\begin{enumerate}
    \item Take the average of the last $m$ iterations and take the averaged parameter values as the final estimates; denote such estimates by $\hat\Theta^{(\text{it})}(m)$, and then its asymptotic variance is (see Proposition 5 in \cite{nielsen2000stochastic})
    \begin{align}
        I(\Theta_0)^{-1} &+ \frac{1}{m}I(\Theta_0)^{-1}[I_p - (I_p + F(\Theta_0))^{-1}] \nonumber\\
        \label{eq: avg-iter-variance}
        &+ \frac{1}{m}I(\Theta_0)^{-1}[I_p - (I_p + F(\Theta_0))^{-1}]F(\Theta_0)(1-F(\Theta_0))^{-1} + o(\frac{1}{m}).
    \end{align}
    \item Take $m$ independent runs of the stochastic EM algorithm and take the average of the $m$ estimates produced by each run; denote such estimates by $\hat\Theta^{\text{(ind)}}(m)$, and then its asymptotic variance is (see Section 4.2.2 in \cite{nielsen2000stochastic})
    \begin{equation}
        \label{eq: avg-chains-variance}
        I(\Theta_0)^{-1} + \frac{1}{m}I(\Theta_0)^{-1}[I_p - (I_p + F(\Theta_0))^{-1}].
    \end{equation}
\end{enumerate}

For instance, if we run the stochastic-EM procedure $m=10$ times in parallel and take the average of the estimates produced by those $10$ independent runs, then the asymptotic variance of these estimates is bounded above by $(1+\frac{1}{10}\times 0.5)I(\Theta_0)^{-1} = 1.05I(\Theta_0)^{-1}$. We can further plug in the estimates 
$\hat\Theta^{\text{(ind)}}(10)$ for $\Theta_0$ to get a variance estimate of $1.05I\left(\hat\Theta^{\text{(ind)}}(10)\right)^{-1}$. In the simulations in Section~\ref{sec:sim-partial} these approximations to the variance are conservative, but not for all parameters. While the Wald-type intervals for most parameters cover the truth nearly 100\% of the time, the $\beta$ and $\eta$ in our simulations are covered 93\% and 81\% of the time, respectively.

Because the marginal likelihood of observed data isn't available, the observed data information matrix $I(\cdot)$
is not immediately obtainable. However, we can nonetheless estimate $I(\cdot)$
 using the Louis identity \citep{louis1982finding}
\begin{equation}\label{eq:louis}
    -\frac{\partial^2 \ell_{\text{obs}}}{\partial \Theta^2} = \mathbb{E}_{\Theta}\left(-\frac{\partial^2 \ell}{\partial \Theta^2}\right) - \text{cov}_{\Theta}\left(\frac{\partial \ell}{\partial \Theta}\right),
\end{equation}
where $\ell_{\text{obs}}$ is the (marginal) log-likelihood of the observed data, and $\ell$ is the log-likelihood of the complete data. Both terms of the right-hand side only involve the complete data likelihood and can be estimated via Monte Carlo approximation using the augmented data samples generated across the iterations of our inference procedure \citep{diebolt1995stochastic}. That is, these samples are already available from the estimation process, and so evaluating Eq.~\eqref{eq:louis} does not require additional sampling.

\section{Simulation experiments}
\label{sec: simulation-experiments}
We now turn to validate the methods of inference via simulation experiments. Synthetic data are simulated using the Gillespie algorithm \citep{gillespie1977exact} according to the generative process described in Section \ref{sec: model-framework}. A generated complete dataset includes the initial contact network $\mathcal{G}_0$, the initial exposed/infectious individual(s) $I_0$, and the full event sequence $\{e_{\ell}\}$, where each event $e_{\ell}$ consists of the event time, the identities of the individuals involved in this event, as well as the event type (exposure, manifestation, recovery, link activation and termination). In addition, we assume that for each person $i$ in the target population, we observe a vector $x_i$ of covariates, which in simulation experiments are randomly sampled binary and standard normal variables. 

The full set of parameters to estimate is $\Theta = \{\beta, b_S, \eta, \varphi, \gamma, \boldsymbol{\alpha}, \boldsymbol\omega\}$. For simplicity, throughout this section, we use the following ground-truth setting for the link rates:
\begin{align*}
    \boldsymbol{\alpha}^T &= (\alpha_{HH0}, \alpha_{HI0}, \alpha_{II0}, \alpha_{HH1}, \alpha_{HI1}, \alpha_{II1})
    = (6, 6, 6, 6, 2, 6)\times 10^{-4};\\
    \boldsymbol{\omega}^T &= (\omega_{HH0}, \omega_{HI0}, \omega_{II0}, \omega_{HH1}, \omega_{HI1}, \omega_{II1})
    = (5, 5, 5, 5, 50, 5)\times 10^{-3}.
\end{align*}
That is, we assume that in the second social behavior stage $\mathcal{T}_1$, $\alpha_{HI}$ (link activation for $H-I$ pair) is reduced and $\omega_{HI}$ (link termination for $H-I$ pair) is increased, mimicking a ``quarantine'' or ``lockdown'' phase. The initial network $\mathcal{G}_0$ is a random Erdős–Rényi graph with edge density $0.05$. Parameters for the epidemic process are chosen as follows:
\begin{equation*}
    \beta = 0.2, \, \eta = 0.2, \, \gamma = 0.1, \, \varphi = 0.2, \, p_s = 0.6.
\end{equation*}

These values are chosen to ensure a low probability of epidemic extinction at the beginning and so that a fair proportion (at least 50\%) of the population has been exposed by the end of the outbreak. 

\subsection{Inference with complete data}
We first validate our iterative inference procedure for complete data, as derived in Section \ref{sec: inference-complete}. Table~\ref{tab: MLE-estimates} presents the mean absolute errors (MAEs), variances and mean square errors (MSEs) for all model parameters across 40 simulations on $N=200$-sized populations. Here we present results for $e^{\eta}$ (true value $\approx 1.22$) instead of $\eta$ as $e^{\eta}$ is the quantity that we directly solve for in inference. In the simulation results presented below, the regression coefficient $b_S$ is set as $(1,1)^T$.  

We can see from Table~\ref{tab: MLE-estimates} that when the complete event sequence is available, the derived inference method can estimate model parameters quite accurately. The most challenging parameters to estimate are $\eta$ (shown in terms of $e^{\eta}$ in the table) and $b_S$, largely because we have to resort to numerical optimization procedures to solve for their MLEs. 


\begin{table}[ht]
\caption{MLE estimates on complete data. For each parameter, we present the mean absolute error (MAE), variance and mean square error (MSE) across 40 independent simulations. For multi-dimensional parameters ($b_S$, $\boldsymbol\alpha$ and $\boldsymbol\omega$), the metrics are averaged over all entries. }
\label{tab: MLE-estimates}
\centering
\begin{tabular}{llll}
  \hline
\textbf{Parameter} & \textbf{MAE} & \textbf{Variance} & \textbf{MSE} \\ 
  \hline
$\beta$ & 0.0359 & 0.0021 & 0.0042 \\ 
  $e^{\eta}$ & 0.2799 & 0.1217 & 0.2493 \\ 
  $b_S$ & 0.2213 & 0.0721 & 0.1654 \\ 
  $\gamma$ & 0.0071 & 8.17$\times 10^{-5}$ & 1.77 $\times 10^{-4}$ \\ 
  $p_s$ & 0.0353 & 0.0013 & 0.0030 \\ 
  $\varphi$ & 0.0158 & 4.80$\times 10^{-4}$ & 0.0010 \\ 
  $\boldsymbol\alpha$ & 0.0007 & 4.13$\times 10^{-7}$ & 1.17$\times 10^{-6}$ \\ 
  $\boldsymbol\omega$ & 0.0086 & 6.64$\times 10^{-6}$ & 2.18$\times 10^{-4}$\\ 
   \hline
\end{tabular}
\end{table}

\subsection{Inference with partial observations}
\label{sec:sim-partial}
Our inference method can accommodate two types of missingness in epidemic observations: (1) exposure times and (2) recovery times. Between these two types of missingness, the former is more difficult to handle as it requires the conditional sampling step derived in Section~\ref{sec: missing-expo-time}. 

First, we test out the central part of our inference procedure. To do this, we hold out \emph{all} the exposure times (i.e., every $t_i^{(E)}$ for each person $i$ who ever got infected) from each simulated dataset and treat those time points as unobserved while considering all other information as observed. We then run the proposed stochastic EM algorithm (without the ``sampling recovery times'' step) on each partial dataset. Columns 2-4 in Table~\ref{tab: miss-expo-estimates} (under ``Missing expo. times'')  summarize the estimation results for some of the model parameters across 40 independent simulations. Same as in Table~\ref{tab: MLE-estimates}, we provide the MAE, variance and MSE for the estimates. 

Next, on the same simulated data, we hold out \emph{all} the recovery times (i.e., $t_i^{(R)}$ for each person $i$ who ever became infectious) and treat both $t_i^{(E)}$ and $t_i^{(R)}$ for each infected person $i$ as unobserved. Now the full stochastic EM inference algorithm is applied to each partial dataset, and the results are summarized in columns 5-7 in Table~\ref{tab: miss-expo-estimates} (under ``Missing both'') . 

As one might expect, compared to inference based on complete data, there is some decrease in accuracy, in particular for $e^{\eta}$, $b_S$ and $\beta$. This decrease in performance is largely due to the variability in the sequential samples of individual local neighborhoods. That is, when the exposure time $t_i^{(E)}$ is unknown and has to be conditionally sampled, the local neighborhood structure for each person $i$ (i.e., $I_i^a(t_i^{(E)})$ and $I_i^s(t_i^{(E)})$) is also unknown and fluctuates throughout the rest of the sampler. This would greatly impact the numerical optimization procedure for these three parameters, as they either directly depend on $I_i^a(t_i^{(E)})$ and $I_i^s(t_i^{(E)})$ or involve the function
\begin{equation*}
    F_i(e^{\eta}) = \int_0^{t_i^{(E)}} (I^a_i(t) +I^s_i(t) e^{\eta})dt,
\end{equation*}
which changes whenever $t_i^{(E)}$ gets updated. Moreover, when the exposure times are unknown, the accumulated amount of infection forces exerted on each susceptible person is also unavailable, which would make solving for $\beta$ and $b_S$ more challenging (see Section S2 of the Supplementary Material \citep{bu2021supplement}
for details on the numerical optimization).


We also note that when recovery times are missing in addition to exposure times, there tends to be more variability in the parameter estimates, since more missingness in the data should tend to induce increased uncertainty. Much of the additional uncertainty is reflected in the estimation of exposure- and latency-related parameters (e.g., $e^{\eta}$, $b_S$ and $\varphi$), as the conditional samplers for exposure times and recovery times are co-dependent, and the complexity in estimating those parameters is more vulnerable to the loss of more information. 


\begin{table}[ht]
\caption{Performance of the stochastic EM inference procedure for simulated datasets with all exposure times missing (columns 2-4) and will both exposure and recovery times missing. The estimate of each parameter is obtain by averaging over the last 20 iterations of the parameter sample chain produced by running the inference algorithm. For each missingness scenario, we present the MAE, variance and MSE for the parameter estimates across 40 independent simulations. }
\label{tab: miss-expo-estimates}
\centering
\begin{tabular}{l|lll|lll}
  \hline
   \multirow{2}{*}{\textbf{Parameter}}
    & \multicolumn{3}{c|}{Missing expo. times}& \multicolumn{3}{c}{Missing both}\\
   \cline{2-7}
 & \textbf{MAE} & \textbf{Variance} & \textbf{MSE} & \textbf{MAE} & \textbf{Variance} & \textbf{MSE} \\ 
  \hline
$\beta$ & 0.0483 & 0.0035 & 0.0069 & 0.0590 & 0.0434 & 0.0581 \\ 
 $e^{\eta}$ & 0.3035 & 0.6719 & 0.2635 & 0.3109 & 0.1976 & 0.3093 \\ 
  $b_S$ & 0.3493 & 0.2038 & 0.3477 & 0.3697 & 0.1549 & 0.3693 \\ 
  $\gamma$ & 0.0061 & 5.61$\times 10^{-5}$ & 1.15$\times 10^{-4}$ & 0.0088 & 1.16$\times 10^{-4}$ & 2.408$\times 10^{-4}$  \\ 
  $p_s$ & 0.0479 & 0.0021 & 0.0045 & 0.0353 & 0.0013 & 0.0030 \\ 
  $\varphi$ & 0.0136 & 0.0003 & 0.0005 & 0.0197 & 0.0003 & 0.0009 \\ 
  $\boldsymbol\alpha$ & 0.0005 & 2.50$\times 10^{-8}$ & 3.37$\times 10^{-7}$ & 0.0006 & 4.76$\times 10^{-8}$ & 5.33$\times 10^{-7}$ \\ 
  $\boldsymbol\omega$ & 0.0088 & 2.00$\times 10^{-6}$ & 2.22$\times 10^{-4}$ & 0.0089 & 7.58$\times 10^{-6}$ & 2.70$\times 10^{-4}$ \\ 
   \hline
\end{tabular}
\end{table}

\begin{figure}[ht]
    \centering
    \includegraphics[width = 0.65\textwidth]{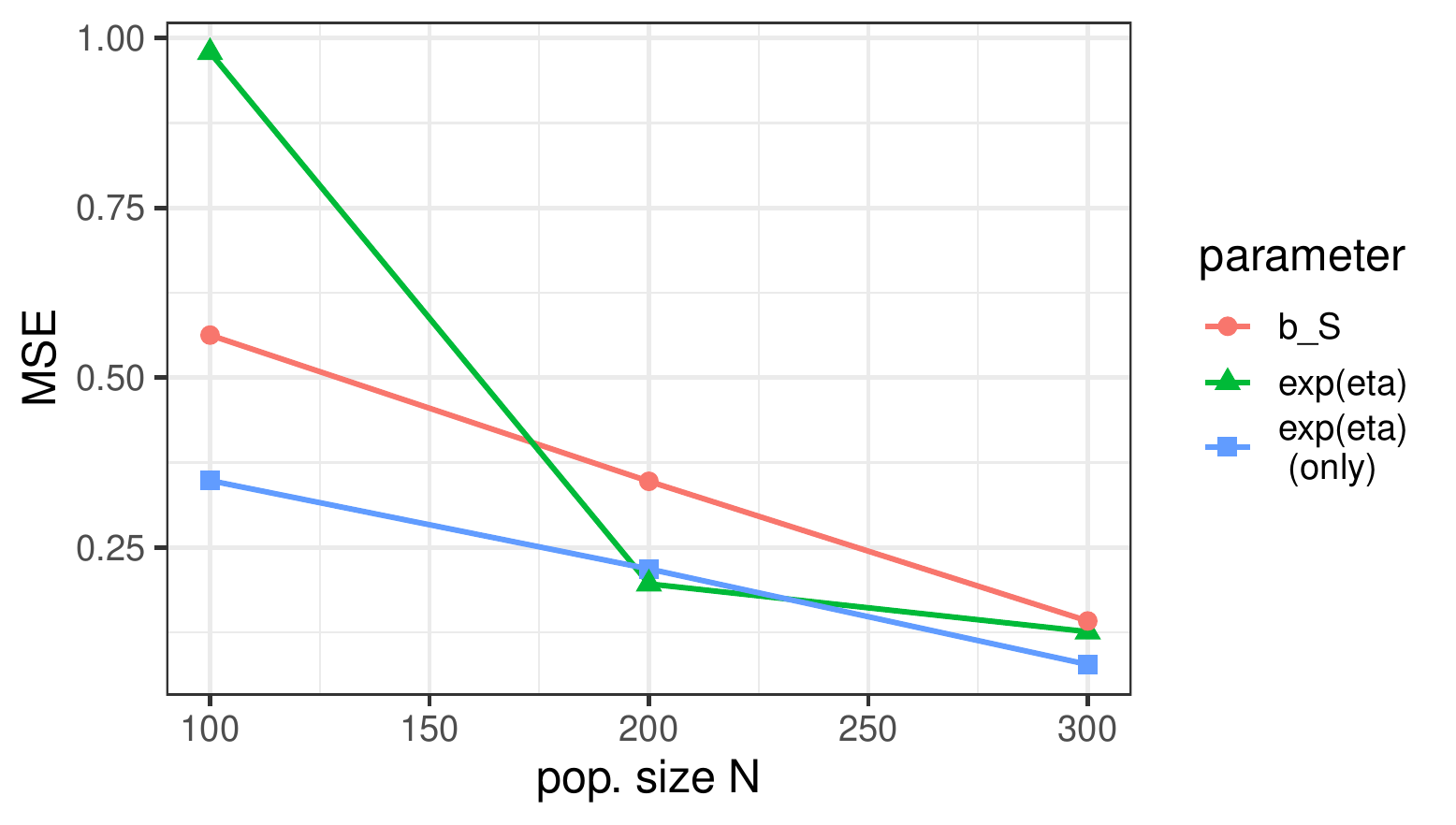}
    \caption{The MSEs for estimating $e^{\eta}$ and $b_S$ with missing exposure and recovery times for different population sizes ($N=100,200,300$). Here, when running Step 3 of the inference algorithm, we fix all other parameters at the true values and only estimate $e^{\eta}$ and $b_S$ (shown in red circles and green triangles); moreover, we run similar estimation procedures but only focus on $e^{\eta}$ (i.e., $b_S$ is fixed at the truth, shown in blue squares). It is clear that when the population size increases and more events are observed, $e^{\eta}$ and $b_S$ are estimated more accurately.}
    \label{fig:eta_bS}
\end{figure}

We further investigate the estimation of these more difficult parameters and note that 
estimation accuracy increases when more data are available. We demonstrate this by conducting simulation experiments for different population sizes ($N=100, 200$ and $300$). Since the number of epidemic events increases approximately linearly in $N$, more events can be observed with a larger population size. For each simulated dataset, we take out \emph{both} the exposure and recovery times and run the inference algorithm, but in Step 3 (solving for the MLEs) we fix all other parameters at the true values and only estimate $e^{\eta}$ and $b_S$. Moreover, we run a separate set of experiments where we only estimate $e^{\eta}$ (i.e., fixing $b_S$ at the truth as well). In Figure~\ref{fig:eta_bS} we present the MSEs of the produced estimates for $e^{\eta}$ and $b_S$ when fixing other parameters (shown as ``b\_S'' and ``exp(eta)'' in red circles and green triangles), and for $e^{\eta}$ only while fixing all other parameters (shown as ``exp(eta) only'' in blue squares). We can see that with increased population size, which yields more observed events, the error in estimating these parameters decreases.

\section{Case study: flu season on a university campus}
\label{sec: case-study}


We now return to the study of transmission of influenza-like illnesses among students on a university campus, where high-resolution contact tracing was conducted to track physical proximity between study subjects. 

This dataset was collected over a 10-week epidemiological study, eX-FLU \citep{aiello2016design}, where inter-personal physical contacts of study participants were surveyed to investigate the effect of social intervention on respiratory infection transmissions. 590 university students enrolled in the study and were asked to respond to weekly surveys on influenza-like illness symptoms and social interactions; they also completed a comprehensive entry survey about demographic information, lifestyles, immunization history, health-related habits, and tendencies of behavioral changes during a flu season or a hypothetical pandemic. 103 individuals among the study population were further recruited to participate in a sub-study in which each study subject was provided a smartphone equipped with an application, iEpi. This application pairs smartphones with other nearby study devices via Bluetooth and thus can record individual-level contacts (i.e., physical proximity) at five-minute intervals.

The iEpi sub-study took place from January 28, 2013 to April 15, 2013 (that is, from week 2 until after week 10 in the main study). Between weeks 6 and 7, there was a one-week spring break (March 1 to March 7), during which epidemic data collection was paused and volume of recorded contacts also dropped considerably. In our application case study, we use data obtained on the $N=103$ sub-study population from January 28 to April 4 (week 2 to week 10), and treat the two periods before and after the spring break as two different social behavior phases. That is, we regard weeks 2-6 as $\mathcal{T}_0$ and weeks 7-10 as $\mathcal{T}_1$ in our analysis. 

Furthermore, we consider two types of ``infectious'' (status $I$) members within the study population: (1) \emph{multi-symptomatic} and (2) \emph{uni-symptomatic}. To maintain notation consistency, we label the former by $I_s$ and the latter by $I_a$. They are defined as follows.
\begin{enumerate}
    \item \emph{multi-symptomatic} ($I_s$), a case with a cough AND one of these three symptoms: fever or feverishness, chills, or body aches.\footnote{This is the definition of ``influenza-like-illnesses''.}
    \item \emph{uni-symptomatic} ($I_a$), a case with a cough, which is an important symptom for influenza. 
\end{enumerate}

For each infection case, we set the reported symptom onset time as the manifestation time (denoted by $t_i^{(I)}$ in previous sections), and treat the exposure time ($t_i^{(E)}$) and recovery time ($t_i^{(R)}$) as unobserved. Since $t_i^{(E)} < t_i^{(I)}$ (as dictated by the design of the assumed epidemic mechanism), we set the plausible latency interval as $L_i = (0, t_i^{(I)})$
. Using weekly surveys (which asked each participant if they felt sick in the past week), we know that the missing recovery times must lie within a 7-day interval for each individual, where the lower and upper bounds are the start and end of a week. Moreover, we assume that all the contact network events are fully observed, as the high-resolution contact tracing can provide timepoints of initiation and termination of all individual-level contacts.\footnote{The timepoints of link activation and termination events are obtained from processing the Bluetooth signals that indicate close proximity of smartphones equipped to the study participants. Technical details in processing the Bluetooth signals are provided in Supplement S6.1 in \cite{bu2020likelihood}.} This suggests that the proposed inference procedure in Section \ref{sec: sto-EM} is applicable to this dataset.

In the rest of this section, we first address a realistic concern of possible external infection sources for the sub-study population, and then present details and discuss results of our data analysis.

\subsection{Inference with external infection sources}
\label{sec: external-infection}
Since the 103 individuals in the dataset are sub-sampled from the 590 study participants, which are also sub-sampled from the entire university campus population, we have to treat the data as observed in an open population instead of a closed one. Therefore, some slight modifications should be made to the model. Specifically, individuals in our target population may get infected by people who are outside of the $N=103$ small population, and we call those people ``external infectors''. 

For simplicity, we represent the joint forces of all external infectors by a single infector that exists outside of the population and exhibits a constant level of transmissibility over time, and this external force of infection is exerted uniformly on all members of the target population. 

For each susceptible individual $j$, let the rate of disease onset (i.e., manifestation) due to external infectors be $\xi_j$, and let this onset rate depend on individual characteristics $x_j$, similar to our treatment of the internal exposure rate $\beta_{ij}$:
\begin{equation}
    \label{eq: external-infection-rate}
    \log \xi_j = \log \xi + x_j^T b_E,
\end{equation}
where $\xi$ denotes the population average external onset rate, and coefficients $b_E$ represent the effects of individual characteristics $x_j$ on subject $j$'s deviation of susceptibility from the average level.

Here $\xi_j$ is the rate of moving from status $S$ directly to either $I_a$ or $I_s$, rather than from $S$ to $E$, and that's why we are naming it the ``external onset rate'' instead of ``external exposure/infection rate''. We are \emph{not} introducing both an exposure rate (like $\beta_{ij}$) and a manifestation rate (like $\varphi$) for external infection cases because of identifiability concerns: since all susceptible people are exposed to the same external infector with time-invariant transmissibility, the exposure rate and manifestation rate would not be identifiable at the same time when the exposure times are not observed. Thus, to ensure identification, we choose to include only one rate instead of two, and the ``onset rate'' can be thought of as the rate of any susceptible individual developing contagiousness due to external infection forces.

Now the set of parameters is extended to $\tilde{\Theta} = \{\beta, \varphi, \gamma, \eta, b_S, \xi, b_E, \boldsymbol\alpha, \boldsymbol\omega\}$, and we can write down a complete data likelihood by slightly modifying Eq.~\eqref{eq: complete-data-likelihood}, where the term related to the new parameters $\xi$ and $b_E$ are separate from the other terms. This means that introducing external cases wouldn't affect parameter estimation of the other parameters at all, and that we can still use the partial data inference procedure detailed in Section~\ref{sec: sto-EM} to analyze the real data. We include details on complete data inference with external cases in Section S4.1 of the Supplementary Material \citep{bu2021supplement}.

\subsection{Data analysis}
\label{sec: data-analysis}

Before applying our model framework to the data, we first discuss how we identify internal and external infection cases and describe the individual characteristics used in the analysis. 
We adopt the following labeling criteria for internal and external infection cases: if an infected person had any infectious contact (within the 103-person population) up to 2 weeks prior to symptom onset, then we label this case as ``internal'', and otherwise this case is labelled as ``external''. This procedure gives us 18 internal cases and 16 external cases in total. Moreover, among all 34 cases, 13 are multi-symptomatic ($I_s$) and 21 are uni-symptomatic ($I_a$). We provide a summary of the breakdown of all infection cases in Table~\ref{tab:case-types}. 
\begin{table}[H]
    \caption{Summary of infection case in the iEpi sub-study data. }
    \label{tab:case-types}
    \centering
    \begin{tabular}{l|rrr}
    \hline
         & $I_s$ & $I_a$ & Total  \\
    \hline
     Internal   &  8 & 10 & 18 \\
     External & 5 & 11 & 16 \\[2pt]
     Total & 13 & 21 & 34 \\
     \hline
    \end{tabular}
\end{table}

We consider the following four individual-level characteristics (all collected from the entry survey) that have previously been linked to disease transmission risk:\footnote{We have included the original survey questions used to calculate the derived covariates ``change\_behavior'' and ``prevention''  in Sections S4.2 and S4.3 of the Supplementary Material \citep{bu2021supplement}.}
\begin{enumerate}
    \item \textbf{flushot}: a binary indicator of whether or not the study subject has taken a flu shot for this year.
    \item \textbf{wash\_opt}: a binary indicator of whether or not the study subject's hand-washing habit is considered ``optimal'', which is derived from survey questions about how long and how frequently one usually washes their hands.
    \item \textbf{change\_behavior}: a derived score that measures how willingly the study subject would change their lifestyle during a hypothetical pandemic; this is calculated by standardizing the numeric sum of the 0/1 scores of 13 Yes/No questions (Yes=1, No=0) about voluntary behavioral changes that essentially translate to reduced social activities or isolation in a lockdown; a higher score represents more willingness in changing one's lifestyle in response to a pandemic. 
    \item \textbf{prevention}: a derived score that measures one's belief in the effectiveness of different preventative practices in reducing the risk of catching the flu; this is calculated by standardizing the numeric sum of response scores (ranging from 1 to 5, 1=strongly disagree, 5=strongly agree) to 6 questions regarding potential preventative measures against flu transmission; a high score represents stronger belief in the effectiveness of preventative practices. 
\end{enumerate}

We perform 20 independent runs of the stochastic-EM inference procedure on the dataset, each time with a different random initialization and 60 burn-in steps. For each run, we take the average of the last 20 iterations (after burn-in) and then average over the 20 averages (across runs) to produce estimates of the parameters. Asymptotic standard errors are obtained using the method described in Section \ref{sec: variance-estimation}; here we obtain a conservative estimate of standard errors by setting $m=20$ and upper-bounding the asymptotic variance matrix by $1.025 I(\hat{\tilde{\Theta}})^{-1}$, where $\hat{\tilde{\Theta}}$ are the final parameter estimates produced by averaging. 

Tables \ref{tab:epi-params-real-data} and \ref{tab:coefs-real-data} present estimates of select parameters of interest. Note that here we take one day as $1$ unit of time. From the epidemic parameter estimates, we can see that for this population, the baseline exposure rate is quite high, indicating fast disease exposure upon contact (it takes approximately 0.22 days on average for an $H-I$ contact to lead to infection if the susceptible individual is not vaccinated and does not wash hands properly); the latency period lasts slightly less than 5 days on average, while recovery from symptoms and contagiousness takes about 6 days on average. The total external infection force experienced by the entire $N=103$-person population is on the scale of $0.00445 \times 103 \approx 0.458$, indicating that on average there would be a disease onset due to external sources every other day if nobody in the study population had a flu shot or washed their hands optimally. In terms of the effects of individual-level covariates, we note that the estimates are associated with relatively large standard errors (indicated in the parentheses), and this is potentially due to the small sample size (in particular, the limited number of infection cases). Nevertheless, the effect of hand-washing (``wash\_opt'') seems to be considerable, given that there is a $11$-fold reduction ($1/e^{-2.42}\approx 11.2$) in the exposure risk if one washes their hands optimally compared to suboptimal hand-washing; this effect appears significant, in that the 95\% Wald-type confidence interval constructed using the conservative standard error estimate is $(-4.054,-0.786)$, which does not cover zero. 

In Table~\ref{tab:net-rates-real-data} we include estimates of several parameters related to the contact network process. Here we emphasize the difference between the change rates of $H-H$ (healthy-healthy) links and $H-I$ (healthy-ill) links, as well as the difference between the two phases ($\mathcal{T}_0$ before spring break and $\mathcal{T}_1$ after). We can clearly see that the link deletion rates for $H-I$ links are higher than those of $H-H$ links in both phases, suggesting that the duration of contact between a healthy person and an infectious person is on average shorter than the contact between two healthy people, probably because the students were cutting meetings short with peers who seemed sick in order to avoid getting infected during the flu season. Moreover, we can clearly see that the level of network activity is much higher (both in terms of establishing and breaking contact) in $\mathcal{T}_0$ (weeks 2 to 6, before spring break)  compared to $\mathcal{T}_1$ (weeks 7 to 10, after spring break) when we compare the rates for phase  $\mathcal{T}_0$ and phase $\mathcal{T}_1$. Such findings are enabled by our model design which allows for different levels of network activities by introducing different time phases.

\begin{table}[H]
    \caption{Estimates of key epidemic parameters, with conservative estimates of asymptotic standard errors.}
    \label{tab:epi-params-real-data}
    \centering
    \begin{tabular}{lrr}
    \hline
    Parameter & Estimate & Standard error \\
    \hline
    $\beta$ (internal exposure) & 4.497 & 2.005  \\
    $\xi$  (external onset)  & 0.00445 & 0.00114 \\
    $\varphi$ (latency) & 0.221 & 0.0591 \\
    $\gamma$ (recovery) & 0.161 & 0.0279\\
    $e^{\eta}$ ($I_s$ v.s. $I_a$ infectiousness) & 0.0622 & 0.0526\\
    $p_s$ (proportion of $I_s$) & 0.382 & 0.0854 \\
    \hline
    \end{tabular}
\end{table}

\begin{table}[H]
    \caption{Estimates of epidemic coefficients on individual characteristics, with conservative asymptotic standard deviations in the parentheses.}
    \label{tab:coefs-real-data}
    \centering
    \begin{tabular}{lrrrr}
    \hline
     & (flushot) & (wash\_opt) & (change\_behavior) & (prevention) \\
     \hline
     $b_S$ (internal exposure) & -0.105 (0.671) & \textbf{-2.42 (0.817)} & -0.201 (0.326) & -0.0541 (0.273) \\
     $b_E$ (external onset)  & -0.805 (0.597) & -0.139 (0.471) & 0.257 (0.263) & -0.0362 (0.273) \\
     \hline
    \end{tabular}
\end{table}

\begin{table}[H]
    \caption{Estimates of link activation and deletion rates for different link types in the two phases ( $\mathcal{T}_0$ spans from week 2 to week 6, and $\mathcal{T}_1$ from week 7 to week 10).}
    \label{tab:net-rates-real-data}
    \centering
    \begin{tabular}{lrrrr}
    \hline
    Link type & $\mathcal{T}_0$ act. &  $\mathcal{T}_1$ act. & $\mathcal{T}_0$ del.   & $\mathcal{T}_1$ del. \\
    \hline
    $H$-$H$ & 0.0181 & 0.000868 & 11.62 & 5.27 \\
    $H$-$I$ & 0.0153 & 0.000653 & 16.62 & 8.71\\
    \hline
    \end{tabular}
\end{table}

Through our data analysis, we have found that proper hand-washing is significantly associated with reduced risk of flu infection, and that there is a considerable external force of infection for the study population. Moreover, study participants exhibit adaptive contact behavior to flu transmission in that contact between healthy and infectious individuals is less frequent and also lasts less time compared to contact between healthy individuals. Our model has also identified differential inter-personal contact patterns in the two observational periods before and after the school break. These findings are consistent with our intuition and are now quantified statistically within a joint inferential framework.

\section{Discussion} 
In this paper, we present a continuous-time Markov chain model for infectious diseases that accounts for individual-level heterogeneity and a dynamic contact structure. Our proposed model can capture the interplay between the epidemic process and network changes, and, more importantly, can describe heterogeneous susceptibility and transmissibility through individual covariates. To accommodate unobserved exposure times and recovery times in real epidemic data, we develop a data-augmented inference procedure based on the stochastic EM algorithm so that we can make use of the complete data likelihood. We also design an efficient method to conditionally sample missing exposure times that are compatible with observed data and respect the dynamic contact network structure. Experiments show that the developed inference procedure performs well on partial data and is able to uncover notable phenomena from modern epidemic data with high-resolution contact tracing. 

It is important to note that the modeling framework we propose is flexible beyond our choice of underlying compartments. That is, our approach can be easily adapted to incorporate notions of reinfection (by allowing some individuals to reenter the susceptible population) or to distinguish between more than two types of infections. In pursuing generalizations of the methodology, introducing additional parameters requires careful consideration of the uncertainty quantification from the stochastic EM algorithm. Although we are able to show that in our setting, the estimated confidence intervals perform well empirically compared to their nominal coverage, they rely on variance approximation formulas, and it is crucial to conduct similar validations in more complex models.  

Finally, our analysis of the iEpi data provides confirmation of the importance of hand-washing on the reduction of the spread of influenza-like-illness. Unlike previous claims in this area, we are able to measure the actual effect on the transmission rate of a disease in an active population with dynamically changing contact patterns. We hope that this development encourages greater data collection of observational high-frequency individual-level data in this area to gain better understanding of other pharmaceutical and non-pharmaceutical interventions. 
For example, future studies will be able to estimate the effectiveness of vaccination in preventing transmission under different social interaction rates and population densities, as well as validate claims about the efficacy of mask-wearing and active social distancing. Importantly, such data can be collected discretely in closed populations and provide invaluable insight into the deployment of public health interventions \citep{motta2021assessment}.


\section*{Supplementary Material}

\begin{itemize}
    \item[A] Supplementary material including derivations, proofs and additional details of inference and data analysis. (Document attached at the end.)
    \item[B] Code and example synthetic data. \texttt{R} and \texttt{Python} code for inference and simulations; since the real data are proprietary, only example synthetic data are provided. GitHub repository: \url{github.com/fanbu1995/EpiNetHetero}.
\end{itemize}


\bibliographystyle{chicago}
\bibliography{ref}       

\newpage

\renewcommand{\theequation}{S\arabic{equation}}
\setcounter{equation}{0}

\renewcommand{\thefigure}{S\arabic{figure}}
\setcounter{figure}{0}

\renewcommand{\thesection}{S\arabic{section}}
\setcounter{section}{0}

\begin{center}
    \huge 
    Supplement to\\ Likelihood-based Inference for partially observed stochastic epidemics with individual heterogeneity
\end{center}

\section{Additional details of the model framework}
\label{supp: addition-model-framework}
As a supplement to the model framework introduced in Section~2, we may consider individual-level covariates in the change rates of links in the contact network. Specifically, given the status of the process at time $t$, $\mathcal{Z}_t$, if $i$ and $j$ are not in contact, then they initiate contact with rate $\alpha_{ijt}$, where
\begin{equation}
    \label{eq:alpha-ij-regression}
    \log \alpha_{ijt}
    = \left[ \alpha_{A_{it}A_{jt}0}\mathbbm{1}(t \in \mathcal{T}_0) +\alpha_{A_{it}A_{jt}1}\mathbbm{1}(t \in \mathcal{T}_1)\right] + x_i^Tb_{\alpha} + x_j^Tb_{\alpha},
\end{equation}
where 
$\alpha_{ABk}$ represents the baseline link activation rate for link type $A-B$ in phase $\mathcal{T}_k$, and $b_{\alpha}$ are the coefficients that describe the additional effects of individual characteristics on the pairwise link activation rate. Similarly, we can introduce the same regression structure to link termination rate $\omega_{ijt}$,
\begin{equation}
    \label{eq:omega-ij-regression}
    \log \omega_{ijt}
    = \left[ \omega_{A_{it}A_{jt}0}\mathbbm{1}(t \in \mathcal{T}_0) +\omega_{A_{it}A_{jt}1}\mathbbm{1}(t \in \mathcal{T}_1)\right] + x_i^Tb_{\omega} + x_j^Tb_{\omega},
\end{equation}
where $b_{\omega}$ are the coefficients that describe the effects of individual characteristics on the pairwise link termination rate.

With these additional terms added to the link change rates, the complete data inference procedure described in Section~3.1 has to be updated. First of all, the complete data likelihood now becomes
\begin{align}
    & L(\Theta; \text{complete data})  \nonumber \\
   = &\beta^{n_E} \gamma^{n_R} \varphi^{n_I} p_s^{n_{I_s}} (1-p_s)^{n_{I_a}} \prod_{i: i \text{ got exposed}} e^{b_S^T x_i} \left[I^{a}_i(t_i^{(E)})+I^{s}_i(t_i^{(E)}) e^{\eta}\right] \nonumber \\
    \times & \prod_{k=0,1}\prod_{(A,B)\in \mathcal{S}} \left[\left(\alpha_{ABk}\right)^{C_{ABk}} \left(\omega_{ABk}\right)^{D_{ABk}}\right] \prod_{i=1}^N e^{N^c_i x_i^Tb_{\alpha}}e^{N^d_i x_i^Tb_{\omega}}  
    \nonumber \\
    \label{eq: complete-data-likelihood-net-regression}
    \times & \exp\left(-\int_0^T \left[\beta \sum_{i=1}^N e^{b_S^T x_i} \left[I^{a}_i(t)+I^{s}_i(t) e^{\eta}\right]\mathbbm{1}(i\text{ is susceptible at } t) + \gamma (I^a(t)+I^s(t)) + \varphi E(t)\right] dt \right)  \\
    \times &  \exp\left( -\int_0^T \sum_{i,j=1:N, i<j} \left[ \alpha_{A_{it}A_{jt}k} e^{x_i^Tb_{\alpha} + x_j^Tb_{\alpha}} (1-\mathbbm{1}_{i-j}(t)) + \omega_{A_{it}A_{jt}k} e^{x_i^Tb_{\omega} + x_j^Tb_{\omega}} \mathbbm{1}_{i-j}(t)\right] dt \right). \nonumber
\end{align}
Here $N_i^c$ and $N_j^c$ denote the total number of link activation and termination events that $i$ has been involved in, respectively. Moreover, $\mathbbm{1}_{i-j}(t)$ is an indicator of whether or not $i$ and $j$ are connected at time $t$. 

Then we can see that all the inference steps related to the epidemic parameters shall remain unchanged, but we need modified steps to estimate $\alpha, b_{\alpha}, \omega$, and $b_{\omega}$. Take partial derivatives of the log-likelihood with respect to these parameters and set them to zero,
\begin{align}
    \label{eq:alpha-derivative-new}
    \frac{\partial \ell}{\partial \alpha_{ABk}} &= \frac{C_{ABk}}{\alpha_{ABk}} - \sum_{i<j} d_{ABk}^{(ij)} e^{(x_i+x_j)^T b_{\alpha}}\quad (\text{for }k=0,1, (A,B)\in\mathcal{S}),\\
    \label{eq:omega-derivative-new}
    \frac{\partial \ell}{\partial \omega_{ABk}} &= \frac{D_{ABk}}{\omega_{ABk}} - \sum_{i<j} c_{ABk}^{(ij)} e^{(x_i+x_j)^T b_{\omega}}\quad (\text{for }k=0,1, (A,B)\in\mathcal{S}),\\
    \label{eq:b-alpha-derivative}
    \frac{\partial \ell}{\partial b_{\alpha}}&= \sum_{i=1}^N N_i^c x_i - \sum_{k=0,1}\sum_{(A,B)\in \mathcal{S}} \alpha_{ABk} \sum_{i<j} \left[d_{ABk}^{(ij)} e^{(x_i+x_j)^T b_{\alpha}}(x_i+x_j)\right],\\
    \label{eq:b-omega-derivative}
    \frac{\partial \ell}{\partial b_{\omega}}&= \sum_{i=1}^N N_i^d x_i - \sum_{k=0,1}\sum_{(A,B)\in \mathcal{S}} \omega_{ABk} \sum_{i<j} \left[c_{ABk}^{(ij)} e^{(x_i+x_j)^T b_{\omega}}(x_i+x_j)\right],
\end{align}
where $d_{ABk}^{(ij)}$ is the total time $i$ and $j$ spend as a \textit{disconnected} $A$-$B$ type pair during $\mathcal{T}_k$, and $c_{ABk}^{(ij)}$ is the total time $i$ and $j$ spend as a \textit{connected} $A$-$B$ type pair during $\mathcal{T}_k$. 

Then we can employ the following iterative procedure to solve for $\alpha$ and $b_{\alpha}$: repeat until convergence or maximum number of iterations is reached:
\begin{itemize}
        \item update $\alpha_{ABk}$ (for $(A,B)\in \mathcal{S}=\{(H,H), (H,I), (I,I)\}$ and $k=0,1$) with
        \begin{equation}
            \hat\alpha_{ABk} = \frac{C_{ABk}}{\sum_{i<j} d_{ABk}^{(ij)} e^{(x_i+x_j)^T b_{\alpha}}};
        \end{equation}
        \item update $b_{\alpha}$ by solving \begin{align}
            0&= \sum_{i=1}^N N_i^c x_i - \sum_{k=0,1}\sum_{(A,B)\in \mathcal{S}} \alpha_{ABk} \sum_{i<j} \left[d_{ABk}^{(ij)} e^{(x_i+x_j)^T b_{\alpha}}(x_i+x_j)\right]\\
            &= \sum_{i<j}^N N_{ij}^c x_{ij} - \sum_{i < j} \left(\sum_{k=0,1}\sum_{(A,B)\in \mathcal{S}} \alpha_{ABk}d_{ABk}^{(ij)}\right) e^{x_{ij}^T b_{\alpha}}x_{ij},
        \end{align}
        where $N_{ij}^c$ is the total counts of link activation between $i$ and $j$ and $x_{ij} = x_i + x_j$, and then this is equivalent to solving for the linear coefficients of a \emph{Poisson regression model} with individual offset $\log\left(\sum_{k=0,1}\sum_{(A,B)\in \mathcal{S}} \alpha_{ABk}d_{ABk}^{(ij)}\right)$.
\end{itemize}

Estimation for $\omega$ and $b_{\omega}$ can be conducted in almost exactly the same manner as in the steps above for $\alpha$ and $b_{\alpha}$, so we will omit the details here. 

\section{Details of complete data inference}
\label{supp: beta-eta-bS-solver}
Here we provide details on the numerical iterative procedure for solving the MLEs of paramters $\beta, \eta$ and $b_S$ by setting the partial derivatives in equations~(8) - (10) to zero. Since it is equivalent to operate with $e^{\eta}$ instead of $\eta$, we directly estimate $e^{\eta}$ in this procedure. Let 
\begin{equation*}
    F_i(e^{\eta}) = \int_0^T (I^a_i(t) +I^s_i(t) e^{\eta})\mathbbm{1}(i\text{ susceptible at } t) dt
\end{equation*}
be the total amount of ``pathogen exposure'' for person $i$, which is a function of $e^{\eta}$. We run the following steps until convergence or the maximum number of iterations is reached:
\begin{itemize}
    \item update $\beta$ with
        \begin{equation*}
            \hat\beta = \frac{n_E}{\sum_{i=1}^N e^{x_i^Tb_S}F_i(e^{\eta})};
        \end{equation*}
    \item update $b_S$ by solving
    \begin{eqnarray}
            0 &=& \sum_{i: i\text{ got infected}} x_i - \beta\sum_{i=1}^N e^{x_i^T b_S} F_i(e^\eta) x_i \nonumber\\
            \label{eq: b_s-obj}
            &=& \sum_{i=1}^N y_i x_i - \sum_{i=1}^N e^{x_i^T b_S + \log \beta + \log F_i(e^\eta)} x_i,
    \end{eqnarray}
    where in $y_i = \mathbbm{1}(i \text{ ever got exposed})$, and solving for the objective function~\eqref{eq: b_s-obj}  is equivalent to solving for the linear coefficients of a \emph{Poisson regression model} with individual offset $\log \beta + \log F_i(e^\eta)$;
    \item update $e^{\eta}$ by numerically solving \footnote{Any built-in solver offered by computational softwares (e.g, ``\texttt{optim}'' function in \texttt{R}) should work.} 
        \begin{equation}
        \label{eq: eta-obj}
            0 = \sum_{i: i\text{ got exposed}}\frac{I^s_i(t_i^{(E)})}{I^a_i(t_i^{(E)}) +I^s_i(t_i^{(E)}) e^{\eta}} - \beta \sum_{i=1}^N e^{x_i^T b_S} \int_0^{t_i^{(E)}} I^s_i(t) dt.
        \end{equation}
\end{itemize}

\section{Derivations of the rejection sampler for missing exposure times}
\label{supp: sample-expo-times}
Given that the risk $\lambda_i(t)$ is a step function, 
the normalizing constant $C_i(t_{\min}^{i},t_{\max}^{i})$ in Equation~(17) can be explicitly evaluated as (note that $\lambda_i(t) \equiv \lambda_{j}$ on each interval $A_j = (t_{j-1}, t_{j})$)
\begin{align}
    C_i(t_{\min}^{i},t_{\max}^{i}) 
    &= \varphi \exp(- \varphi t_i^{(I)}) \sum_{j=1}^{n_i} \lambda_j \exp\left(-\sum_{k < j} \text{len}(A_k)\lambda_k + \lambda_j t_{j-1}\right) \times  \nonumber\\
    & \hspace{1.4in} (t_j - t_{j-1})^{\mathbbm{1}(\varphi = \lambda_j)} \left(\frac{e^{(\varphi - \lambda_j) t_j} - e^{(\varphi - \lambda_j) t_{j-1}}}{\varphi - \lambda_j}\right)^{\mathbbm{1}(\varphi \neq \lambda_j)}.
\end{align}

If we don't have prior knowledge about the latency period, we should perhaps search for possible exposure time between $0$ and the manifestation time. That is, we can adopt the natural choice of $L_i = [0, t_{i}^{(I)}]$, and then the density (17) is simplified into
\begin{align}
    &p_i(t \mid t_i^{(I)}, \beta, \delta_i, \eta, \varphi, \text{network events}) \nonumber\\
    \label{eq: simp-expo-time-dens}
    =& \frac{\lambda_{i}(t) \exp\left(-\int_{0}^t \lambda_{i}(u)du\right) \times \varphi \exp(- \varphi (t_i^{(I)} - t))\mathbbm{1}(0 < t < t_i^{(I)})}{C_i(0, t_{i}^{(I)})}. 
\end{align}

Let
\begin{equation}
\label{eq:proposal-density-app}
    q_i(t) = \frac{\lambda_{i}(t) \exp\left(-\int_{0}^t \lambda_{i}(u)du\right)\mathbbm{1}(0 < t < t_i^{(I)})}{1-\exp\left(-\int_{0}^{t_i^{(I)}} \lambda_{i}(u)du\right)}
\end{equation}
be the density function of a truncated inhomogeneous Exponential distribution with rate function $\lambda_{i}(t)$, and then we would have
\begin{align}
    \frac{p_i(t)}{q_i(t)} =& \frac{\left[1-\exp\left(-\int_{0}^{t_i^{(I)}} \lambda_{i}(u)du\right)\right]\times \varphi \exp(- \varphi (t_i^{(I)} - t))}{C_i(0, t_{i}^{(I)})} \nonumber\\
    \leq & \frac{\left[1-\exp\left(-\int_{0}^{t_i^{(I)}} \lambda_{i}(u)du\right)\right]\times \varphi \exp(- \varphi (t_i^{(I)} - t_i^{(I)}))}{C_i(0, t_{i}^{(I)})} \nonumber\\
    =& \frac{\exp\left(\varphi t_{i}^{(I)}\right) \int_{0}^{t_i^{(I)}} \lambda_{i}(t) \exp\left(-\int_{0}^t \lambda_{i}(u)du\right) dt}{\int_{0}^{t_i^{(I)}} \exp\left(\varphi t\right) \lambda_{i}(t) \exp\left(-\int_{0}^t \lambda_{i}(u)du\right) dt} 
    =: M, \nonumber
\end{align}
and it's straightforward to see that $M > 1$. 

This
suggests that we can sample exposure time $t_i^{(E)}$ from $p_i(t)$ via \emph{rejection sampling} with proposal density $q_i(t)$, as described in Section~
4.1.


\section{More details of the real data analysis}
\label{supp: details-real-data}

\subsection{Complete data inference with external infection cases}

\begin{figure}[H]
    \centering
    \includegraphics[width=0.9\textwidth,page=7,trim={0 0.8in 0 0.8in},clip]{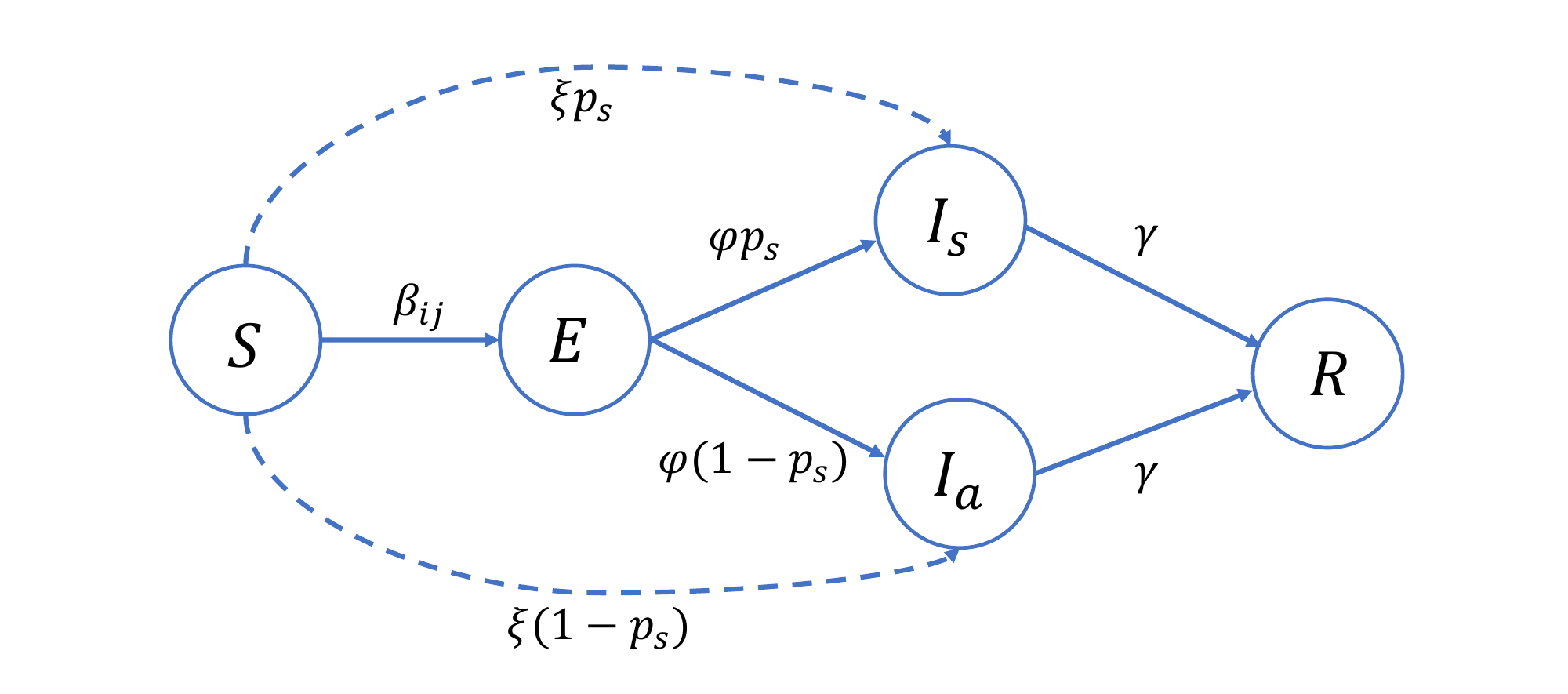}
    \caption{Diagram of the epidemic model with external onset transitions.}
    \label{fig:diagram-external}
\end{figure}

In Figure~\ref{fig:diagram-external}, we provide a diagram of the epidemic model with external onsets. Here external onset is the transition between $S$ and $I$, with the $E$ status subsumed. 

For convenience, when working with the real data, we assume that we have knowledge of which cases are internal infection cases and which are external ones. \footnote{Details on how we identify internal and external cases are discussed in Section \ref{sec: data-analysis}.} Denote the former set of cases by $\mathcal{I}^{(\text{int})}$ and the latter by $\mathcal{I}^{(\text{ext})}$. 

Now the set of parameters are extended to $\tilde{\Theta} = \{\beta, \varphi, \gamma, \eta, b_S, \xi, b_E, \boldsymbol\alpha, \boldsymbol\omega\}$, and
the complete data likelihood 
(7) should be modified into
\begin{align}
    & \tilde{L}(\tilde\Theta; \text{complete data})  \nonumber \\
   = &\beta^{n_E} \gamma^{n_R} \varphi^{n_I^{(\text{int})}} p_s^{n_{I_s}} (1-p_s)^{n_{I_a}} \prod_{i: i \in \mathcal{I}^{(\text{int})}} e^{b_S^T x_i} \left[I^{a}_i(t_i^{(E)})+I^{s}_i(t_i^{(E)}) e^{\eta}\right] \times \xi^{n_I^{(\text{ext})}} \prod_{i: i \in \mathcal{I}^{(\text{ext})}} e^{b_E^T x_i} \nonumber\\
    \times & \prod_{k=0,1}\prod_{(A,B)\in \mathcal{S}} \left[\left(\alpha_{ABk}\right)^{C_{ABk}} \left(\omega_{ABk}\right)^{D_{ABk}}\right] 
    \nonumber \\
    \label{eq: complete-data-likelihood-external}
    \times & \exp\left(-\int_0^T \left[\beta \sum_{i=1}^N e^{b_S^T x_i} \left[I^{a}_i(t)+I^{s}_i(t) e^{\eta}\right]\mathbbm{1}(i\text{ is susceptible at } t) + \gamma (I^a(t)+I^s(t)) + \varphi E(t)\right] dt \right)  \\
    \times & \exp\left(-\xi \sum_{i=1}^N t_i^{(I)}e^{b_E^T x_i} \right) 
    \exp\left( -\int_0^T  \sum_{k=0,1}\sum_{(A,B)\in \mathcal{S}} \left[ \alpha_{ABk} M^d_{AB}(t) + \omega_{ABk} M^c_{AB}(t) \right] \mathbbm{1}(t \in \mathcal{T}_k)dt \right). \nonumber
\end{align}

In addition to the notation explained in Section 
2, here $n_I^{(\text{int})}, n_I^{(\text{ext})}$ denote the total number of infection (manifestation) cases due to internal and external sources, respectively. Again, similar to previous definitions, for convenience, if an individual $i$ never got infected, then we set $t_i^{(E)} = t_i^{(I)} = t_i^{(R)} = T$. 

Examining the above expression, it is obvious that estimation of all previously existing parameters are not really impacted, except that all formulas related to $\beta, b_S, \eta$ and $\varphi$ should be restricted to internal infection cases instead of all infection cases. Therefore, equations 
(12) - (15) shall remain completely unchanged, and the rest of the partial derivatives should be modified slightly, on top of adding two more for $\xi$ and $b_E$:
\begin{align}
    \label{eq:beta-derivative-2}
    \frac{\partial \ell}{\partial \beta} &= \frac{n_E}{\beta} - \sum_{i=1}^N e^{x_i^T b_S} \left[\int_0^{t_i^{(E)}} (I^a_i(t) +I^s_i(t) e^{\eta}) dt\right],\\
    \label{eq:b-S-derivative-2}
    \frac{\partial \ell}{\partial b_S} &= \sum_{i: i\in \mathcal{I}^{(\text{int})}} x_i - \beta\sum_{i=1}^N e^{x_i^T b_S} \left[\int_0^{t_i^{(E)}} (I^a_i(t) +I^s_i(t) e^{\eta})dt\right] x_i,\\
    \label{eq:Eeta-derivative-2}
    \frac{\partial \ell}{\partial e^{\eta}} &= \sum_{i: i\in \mathcal{I}^{(\text{int})}}\frac{I^s_i(t_i^{(E)})}{I^a_i(t_i^{(E)}) +I^s_i(t_i^{(E)}) e^{\eta}} - \beta \sum_{i=1}^N e^{x_i^T b_S} \int_0^{t_i^{(E)}} I^s_i(t) dt,\\
    \label{eq:varphi-derivative-2}
    \frac{\partial \ell}{\partial \varphi} &= \frac{n_I^{(\text{int})}}{\varphi} - \int_0^T E(t) dt,\\
    \label{eq:xi-derivative}
     \frac{\partial \ell}{\partial \xi} &= \frac{n_I^{(\text{ext})}}{\xi} - \sum_{i=1}^N t_i^{(I)}e^{b_E^T x_i},\\
    \label{eq:b-E-derative}
    \frac{\partial \ell}{\partial b_E} &= \sum_{i: i\in \mathcal{I}^{(\text{ext})}} x_i -\xi \left[\sum_{i=1}^N t_i^{(I)}e^{b_E^T x_i}\right] x_i.
\end{align}
Same as described in Section 
3.1, solving for MLEs requires setting all partial derivatives to zero, which leads to an updated inference method that includes all steps in Section 
3.1, with an additional iterative procedure that runs until convergence or the maximum number of iterations is reached:
    \begin{itemize}
        \item update $\xi$ with
        \begin{equation}
            \label{eq:xi-update}
            \hat{\xi} = \frac{n_I^{(\text{ext})}}{\sum_{i=1}^N t_i^{(I)}e^{b_E^T x_i}};
        \end{equation}
        \item update $b_E$ by solving
        \begin{align}
            0 =& \sum_{i: i\in \mathcal{I}^{(\text{ext})}} x_i - \xi\left[\sum_{i=1}^N t_i^{(I)}e^{b_E^T x_i}\right] x_i \nonumber\\
            =& \sum_{i=1}^N z_i x_i - \sum_{i=1}^N e^{x_i^Tb_E + \log \xi + \log t_i^{(\text{I})}} x_i,
        \end{align}
        where $z_i = \mathbbm{1}(i\in \mathcal{I}^{(\text{ext})})$, and (very similar to our treatment to $b_S$   in  \eqref{eq: b_s-obj}) this is equivalent to solving for the linear coefficients of a Poisson regression model with individual offset $\log \xi + \log t_i^{(\text{I})}$. 
    \end{itemize}

Therefore, with this updated inference procedure for maximum likelihood estimation from complete data, we can run the inference algorithm stated in Section 
4 on the real data, where epidemic observations are incomplete. Note that the addition of the external infection sources doesn't affect conditional sampling of the missing exposure times or recovery times, so Step 1 and Step 2 of the inference scheme remains unchanged, while only Step 3 is modified to the procedure discussed above.

\subsection{Survey questions used to derive individual characteristic ``change\_behavior''} The following prompt is copied verbatim from the original eX-FLU survey:

The next section asks questions about a possible outbreak of pandemic flu in the US-a new type of flu that spreads rapidly among humans and causes severe illness. Imagine that a lot of people were getting very sick from the flu and the flu was spreading rapidly from person to person when you answer the following questions. During an outbreak, would you voluntarily make the following changes to your life? (Select Yes (1) or No (0). )
\begin{enumerate}
    \item  Cancel plans with friends, other students, or family members;
    \item  Avoid busy public places-e.g., shopping areas, movie theaters, restaurants;
    \item Cancel inter-state or international travel plans;
    \item  Stock up on food and/or other necessities;
    \item Avoid public transportation, including University buses;
    \item Stay in  your residence hall room if you were feeling sick;
    \item Stay in your residence hall room to avoid contact with sick people;
    \item Reduce contact with people outside of your residence;
    \item Wear a face mask when out in public;
    \item Be absent from class;
    \item Be absent from work;
    \item Assist sick neighbors or friends by bringing them food or supplies;
    \item Limit shopping to living essentials, e.g. food, medicine.
\end{enumerate}

\subsection{Survey questions used to derive individual characteristic ``prevention''}
The following prompt is copied verbatim from the original eX-FLU survey:

Do you believe that the following practices reduce your risk of catching flu? (Score from 1 to 5; $1=$ strongly disgree, $5=$strongly agree.)
\begin{enumerate}
    \item Reducing the number of people you meet over a day;
    \item Avoiding public transportation, including university buses;
    \item Cleaning or disinfecting things you might touch;
    \item Washing your hands regularly with soap and water;
    \item Wearing a face mask when out in public;
    \item Avoiding hospitals or doctors’ offices.
\end{enumerate}

\end{document}